\documentclass[useAMS,usenatbib,usegraphicx]{mn2e}
\usepackage{amsmath}

\title[Young globular clusters in the M31 halo]{Young accreted globular clusters in the outer halo of M31}
\author[A.~D.~Mackey et al.]{A.~D.~Mackey$^{1}$, 
A.~P.~Huxor$^{2}$, A.~M.~N.~Ferguson$^{3}$, M.~J.~Irwin$^{4}$, J.~Veljanoski$^{3}$,\newauthor
A.~W.~McConnachie$^{5}$, R.~A.~Ibata$^{6}$, G.~F.~Lewis$^{7}$ and N.~R.~Tanvir$^{8}$\\
$^{1}$RSAA, The Australian National University,
Mount Stromlo Observatory, Cotter Road, Weston Creek, ACT 2611, Australia \\
$^{2}$Astronomisches Rechen-Institut, Universit\"{a}t Heidelberg, M\"{o}nchhofstra{\ss}e 12-14, 
69120 Heidelberg, Germany \\
$^{3}$Institute for Astronomy, University of Edinburgh, Royal Observatory, Blackford Hill,
Edinburgh, EH9 3HJ, UK \\
$^{4}$Institute of Astronomy, University of Cambridge, Madingley Road, Cambridge, CB3 0HA, UK \\
$^{5}$NRC Herzberg Institute for Astrophysics, 5071 West Saanich Road, Victoria, British Columbia,
Canada V9E 2E7\\
$^{6}$Observatoire Astronomique de Strasbourg, 11 rue de l'Universit\'{e}, 67000 Strasbourg, France \\
$^{7}$Sydney Institute for Astronomy, School of Physics, A28, The University of Sydney, NSW 2006, Australia \\
$^{8}$Department of Physics and Astronomy, University of Leicester, University Road, Leicester,
LE1 7RH, UK
}
\begin{document}

\date{Draft version \today.}

\pagerange{\pageref{firstpage}--\pageref{lastpage}} \pubyear{2012}

\maketitle

\label{firstpage}

\begin{abstract}
We report on Gemini/GMOS observations of two newly discovered globular clusters in the outskirts
of M31. These objects, PAndAS-7 and PAndAS-8, lie at a galactocentric radius of $\approx 87$ kpc and are
projected, with separation $\approx 19$ kpc, onto a field halo substructure known as the South-West Cloud. 
We measure radial velocities for the two clusters which confirm that they are almost certainly physically 
associated with this feature. Colour-magnitude diagrams reveal strikingly short, exclusively red horizontal 
branches in both PA-7 and PA-8; both also have photometric $[$Fe$/$H$] = -1.35 \pm 0.15$. At this 
metallicity, the morphology of the horizontal branch is maximally sensitive to age, and we use the distinctive 
configurations seen in PA-7 and PA-8 to demonstrate that both objects are very likely to be at least 2 Gyr 
younger than the oldest Milky Way globular clusters. Our observations provide strong evidence for young 
globular clusters being accreted into the remote outer regions of M31 in a manner entirely consistent 
with the established picture for the Milky Way, and add credence to the idea that similar processes play a 
central role in determining the composition of globular cluster systems in large spiral galaxies in general.
\end{abstract}


\section{Introduction}
It has long been recognized that some fraction of the globular clusters in the Milky Way halo have very likely
been accumulated via the accretion and destruction of their satellite host galaxies, rather than having formed
{\it in situ}. However much of the observational support for this picture is circumstantial rather than direct. 
In their seminal paper \citet{searle:78} showed that halo clusters outside the solar circle exhibit no abundance 
gradient with Galactocentric radius, and further that the horizontal branch (HB) morphologies for these objects 
(taken as a proxy for age) show little correlation with abundance. They postulated that the outer 
globular clusters did not form during the initial collapse of the Galaxy but originated in smaller ``proto-galactic 
fragments'' that were subsequently accreted into the Galactic potential well, even long after the formation of
the central regions of the Milky Way was complete.

Modern studies with vastly improved datasets still support this scenario. It is now known that a wide variety 
of properties commonly associated with globular clusters in the outer parts of the Milky Way, such as their
ages (measured directly from precise main-sequence turn-off photometry), kinematics, HB morphologies, 
spatial locations, luminosities, and sizes, are consistent with an external origin 
\citep*[see e.g.,][]{zinn:93,mackey:04,mackey:05,marinfranch:09,forbes:10,dotter:10,dotter:11,keller:12}.
Under the assumption that age is the dominant {\it second parameter} controlling horizontal branch morphology, 
\citet{zinn:93} labelled the ensembles of globular clusters with red and blue HBs at given $[$Fe$/$H$]$ as the ``young'' 
and ``old'' halo systems, respectively. That young halo clusters typically have 
ages several Gyr younger than old halo members has now been explicitly demonstrated \citep[e.g.,][]{dotter:10,dotter:11}, 
and it is these objects,
with their characteristic diffuse structures and frequently retrograde orbits, that are the strongest candidates
for having been accreted into the Milky Way.

The ``smoking gun'' for the accretion scenario was the discovery of the disrupting Sagittarius dwarf 
\citep*{ibata:94} and the recognition that at least four remote Galactic globular clusters 
(M54, Arp 2, Terzan 7 and 8) are in fact members of this galaxy caught in the midst of their arrival into the 
Milky Way halo \citep*{ibata:95,dacosta:95}. It has been shown more recently that at least one young halo
globular cluster widely separated from the main body of Sagittarius on the sky, Palomar 12, also
once belonged to that galaxy \citep{md:02,cohen:04}; furthermore, a handful of additional clusters are strong 
candidates for being ex-members \citep*[e.g.,][]{bellazzini:03,law:10,forbes:10}. 

If a significant fraction of remote Milky Way globular clusters have indeed been accreted, it might
be expected that the outskirts of these objects would be an excellent place to search for the remnants of
their host galaxies. However, despite a significant investment of effort \citep[e.g.,][]{sohn:03,md:04} and a 
few tantalising recent hints \citep[e.g.,][]{olszewski:09,sollima:12}, no such streams have yet been definitively 
identified. To date, the handful of 
clusters arriving along with the Sagittarius dwarf represent the only {\it direct} evidence for the accretion scenario; 
all else remains circumstantial. In particular, despite the compelling properties of the young halo ensemble, 
it is not clear to what extent these objects are representative of the entirety of the accreted
population -- several of the Sagittarius clusters are old halo members, and indeed a number of the clusters
seen in the LMC and the Fornax dSph would also be classified as such \citep[e.g.,][]{mackey:04,mackey:05}.

M31 is of vital importance in providing a unique external system for addressing the above problems; it 
is the only other large galaxy in which globular clusters may be resolved into individual stars. Wide-field surveys
conducted by our group, including, most recently, the {\it Pan-Andromeda Archaeological Survey} 
\citep[PAndAS;][]{mcconnachie:09}, have uncovered a wealth of streams and substructure in the M31 field
halo \citep[see also][]{ferguson:02,ibata:07}, along with a large number of remote globular clusters with
projected radii in the range $\approx 15-145$ kpc \citep[][2013, in prep.]{huxor:05,huxor:08}. 
We have previously demonstrated that there is a striking correlation between a number of the tidal streams seen in
the M31 halo and the positions of many of the globular clusters \citep{mackey:10a}; now that the PAndAS
footprint is complete, it has become clear that this association apparently holds across the entire M31 halo
beyond $\approx 30$ kpc (Mackey et al. 2013, in prep.). The unavoidable conclusion from these observations is 
that much of the outer M31 globular cluster system ($\ga 80\%$) has been assembled via the accretion of 
cluster-bearing satellite galaxies, just as has long been suspected in the Milky Way.

We have embarked on a detailed observational programme aimed at measuring (i) radial velocities for, and
(ii) the resolved properties of, remote M31 globular clusters, especially those projected onto tidal streams and overdensities.
These observations will confirm (or refute) individual associations between clusters and substructures, and allow
us to explore in detail the characteristics -- in particular the metallicities, HB morphologies, and structures --
of accreted M31 globular clusters for direct comparison with Milky Way subsystems such as the young and old 
halo ensembles.

\begin{figure*}
\begin{minipage}{175mm}
\begin{center}
\includegraphics[width=160mm]{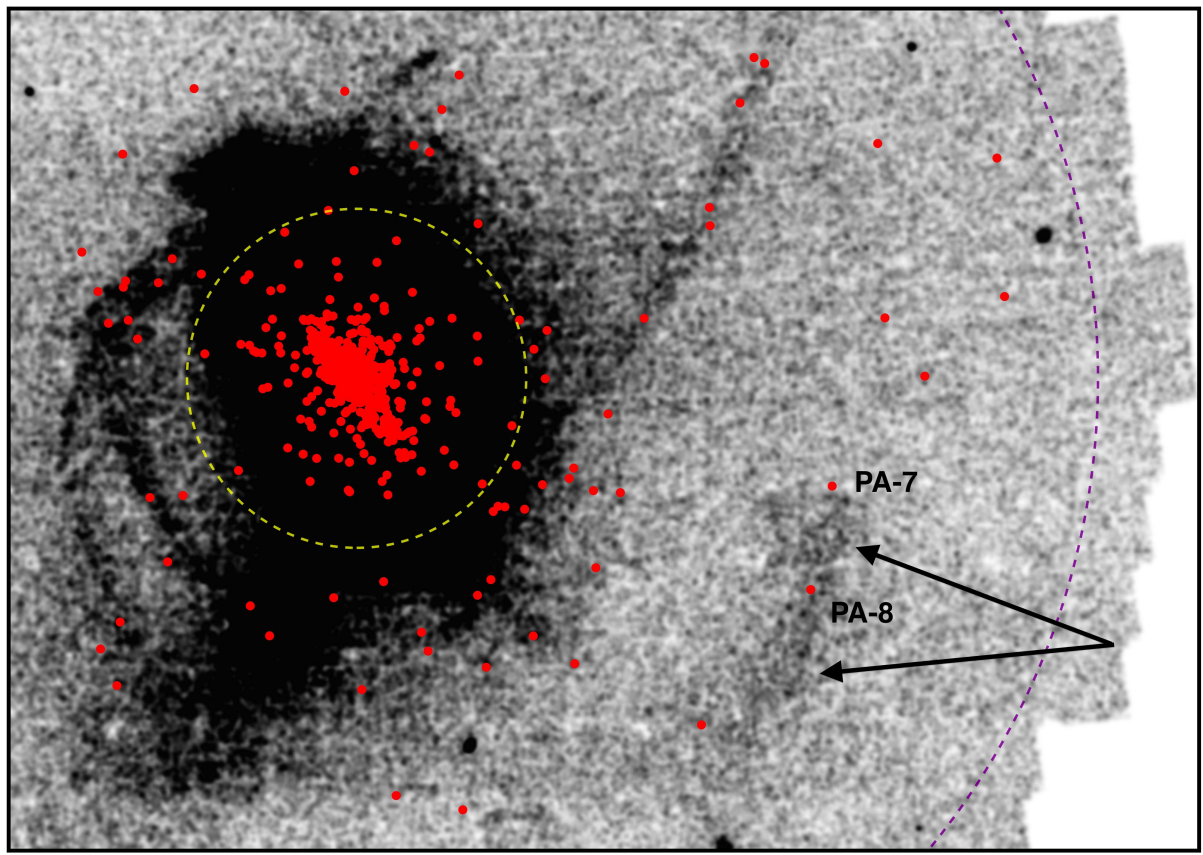}
\caption{Spatial density map, excised from the final PAndAS footprint, of stellar sources possessing luminosities and colours consistent with being metal-poor red giant branch stars ($[$Fe$/$H$] \la -1.4$) in the M31 halo. North is up and east is to the left. The two dashed circles represent projected galactocentric radii of $30$ and $130$ kpc ($2.2$ and $9.6$ degrees). Globular clusters are marked with red points. In the inner regions of M31 these come from the Revised Bologna Catalogue \citep{galleti:04}; outside $\approx 30$ kpc almost all have been discovered from our wide-field surveys, including PAndAS. The two clusters on which we focus in this paper, PA-7 and PA-8, are labelled. These project onto the South-West Cloud, indicated with arrows. A third cluster, not studied here, also sits on this field substructure -- this is PA-14, to the south-east of PA-8. Also notable in the field of view are the narrow North-West Stream, the Giant Stream to the south, and two tangential streams to the east \citep[see][]{ibata:07,mcconnachie:09}.}
\label{f:map}
\end{center}
\end{minipage}
\end{figure*}

In this paper we report on some of the first results of this work. We have targeted two newly-discovered M31
globular clusters from the PAndAS survey, PAndAS-7 and PAndAS-8 (Huxor et al. 2013, in prep.), with imaging
and spectroscopic observations using the GMOS instrument on Gemini North. Assuming the usual 
M31 distance modulus of $(m-M)_0 = 24.47$, PA-7 and PA-8 lie at projected galactocentric radii of $86$ and 
$88$ kpc, respectively, and are separated by a distance of $81.7\arcmin \approx 18.6$\ kpc. Both project onto
a field halo substructure known as the South-West Cloud, as shown in Figure \ref{f:map} \citep[see also][]{mackey:10a}.

So far little is known about either the target clusters or the South-West Cloud. Based on the PAndAS discovery 
images both PA-7 and PA-8 appear sub-luminous, with $M_V \approx -4.9$ and $-5.3$, respectively.
With $(V-I)_0 \sim 1.0$, they are also mildly redder than other globular clusters at similar 
radii \citep[which typically have $(V-I)_0 \approx 0.9$ as in e.g.,][]{huxor:11}. The South-West Cloud was first
described by \citet{mcconnachie:09}, and is one of the more diffuse substructures identified in the M31 halo 
to date. Density maps constructed using different photometric cuts 
\citep[such as Fig. \ref{f:map}, see also][]{richardson:11}, indicate that it must be predominantly composed of 
stars with $[$Fe$/$H$] \la -1$. No kinematics have yet been measured for the South-West Cloud; however, 
\citet{lewis:12} describe a large spur of H{\sc i} gas extending from the central parts of M31 to very nearly 
overlap with this stellar substructure. While any link between the two remains circumstantial, we note that the 
H{\sc i} velocity is $\sim -470$\ km$\,$s$^{-1}$, with width $\approx 200$\ km$\,$s$^{-1}$.

In what follows we describe our data acquisition and reduction (Section \ref{s:obs}), the results inferred
from radial velocity measurements and the cluster colour-magnitude diagrams (Section \ref{s:results}),
and the implications for our understanding of accretion processes in assembling both M31 and the Milky Way
globular cluster systems (Section \ref{s:discuss}).

\section{Observations and data reduction}
\label{s:obs}
\subsection{Imaging and photometry}
We imaged PA-7 and PA-8 over a number of nights in August and September 2008, as indicated in Table
\ref{t:obs}, using the Gemini Multi-Object Spectrograph (GMOS) at the $8.1$m Gemini North telescope on 
Mauna Kea, Hawaii. The data were obtained in queue mode via program GN-2008B-Q-22 (PI: Mackey) during 
clear, photometric conditions and under excellent seeing ($0.4\arcsec - 0.65\arcsec$).

\begin{table*}
\centering
\caption{Log of imaging and spectroscopic observations of the two target clusters and two velocity reference clusters.}
\begin{tabular}{@{}cccccccccc}
\hline \hline
Cluster & \hspace{2mm} & \multicolumn{2}{c}{Coordinates (J2000.0)} & \hspace{2mm} & Filter or & Number of & Total & Dates & Image\\
Name  & & RA & Dec & & Grating & Frames & Exposure & Observed & Quality \\
\hline
PAndAS-7 & & $00\;10\;51.3$ & $+39\;36\;00.0$ & & $g\arcmin$ & $6$ & $3600$s & $2008$-$08$-$11$, $2008$-$09$-$05$ & $0.40-0.60\arcsec$ \\
                 & &                &                & & $i\arcmin$ & $5$ & $2400$s & $2008$-$08$-$11$, $2008$-$08$-$31$ & $0.55-0.65\arcsec$ \\ 
                 & &                &                & & R831 & $6$ & $5400$s & $2010$-$07$-$20$ & $\le 0.75\arcsec$ \\ 
\hline
PAndAS-8 & & $00\;12\;52.4$ & $+38\;17\;48.0$ & & $g\arcmin$ & $6$ & $3600$s & $2008$-$09$-$03$ & $0.50-0.55\arcsec$ \\
                 & &                &                & & $i\arcmin$ & $5$ & $2400$s & $2008$-$09$-$03$ & $0.40-0.50\arcsec$ \\
                 & &                &                & & R831 & $6$ & $5400$s & $2010$-$07$-$20$, $2010$-$08$-$15$ & $\le 0.75\arcsec$ \\ 
\hline
G1             & & $00\;32\;46.5$ & $+39\;34\;40.7$ & & R831 & $9$ & $1350$s & $2010$-$07$-$20$ & $\le 0.75\arcsec$ \\ 
\hline
MGC1        & & $00\;50\;42.5$ & $+32\;54\;58.8$ & & R831 & $9$ & $1800$s & $2011$-$08$-$02$ & $\le 0.75\arcsec$ \\ 
\hline
\label{t:obs}
\end{tabular}
\end{table*}
 
The GMOS imager \citep{hook:04} is comprised of three adjacent $2048\times4096$ pixel CCDs 
separated by gaps of $\sim 2.8\arcsec$, and has a field of view (which does not cover the full 
CCD package) of $5.5\arcmin \times 5.5\arcmin$. To take advantage of the high quality conditions 
we employed unbinned imaging, resulting in a plate scale of $0.0727$ arcsec$/$pixel. 
We obtained our observations in the GMOS $g\arcmin$ and $i\arcmin$ filters, which are 
similar, but not identical, to the $g$ and $i$ filters used by the Sloan Digital Sky Survey (SDSS), and the
$g\arcmin$ and $i\arcmin$ filters employed by the MegaCam imager on the Canada-France-Hawaii
Telescope (CFHT). We took six images with the $g\arcmin$ filter and five with the $i\arcmin$ filter, 
arranged in a $3\times2$ dither pattern with a step size of $5\arcsec$ designed to eliminate the
gaps between the CCDs and provide continuous coverage of our field. Exposure durations were 
$600$s per image for the $g\arcmin$ filter and $480$s per image for the $i\arcmin$ filter.

\begin{figure*}
\begin{minipage}{175mm}
\begin{center}
\includegraphics[width=70mm]{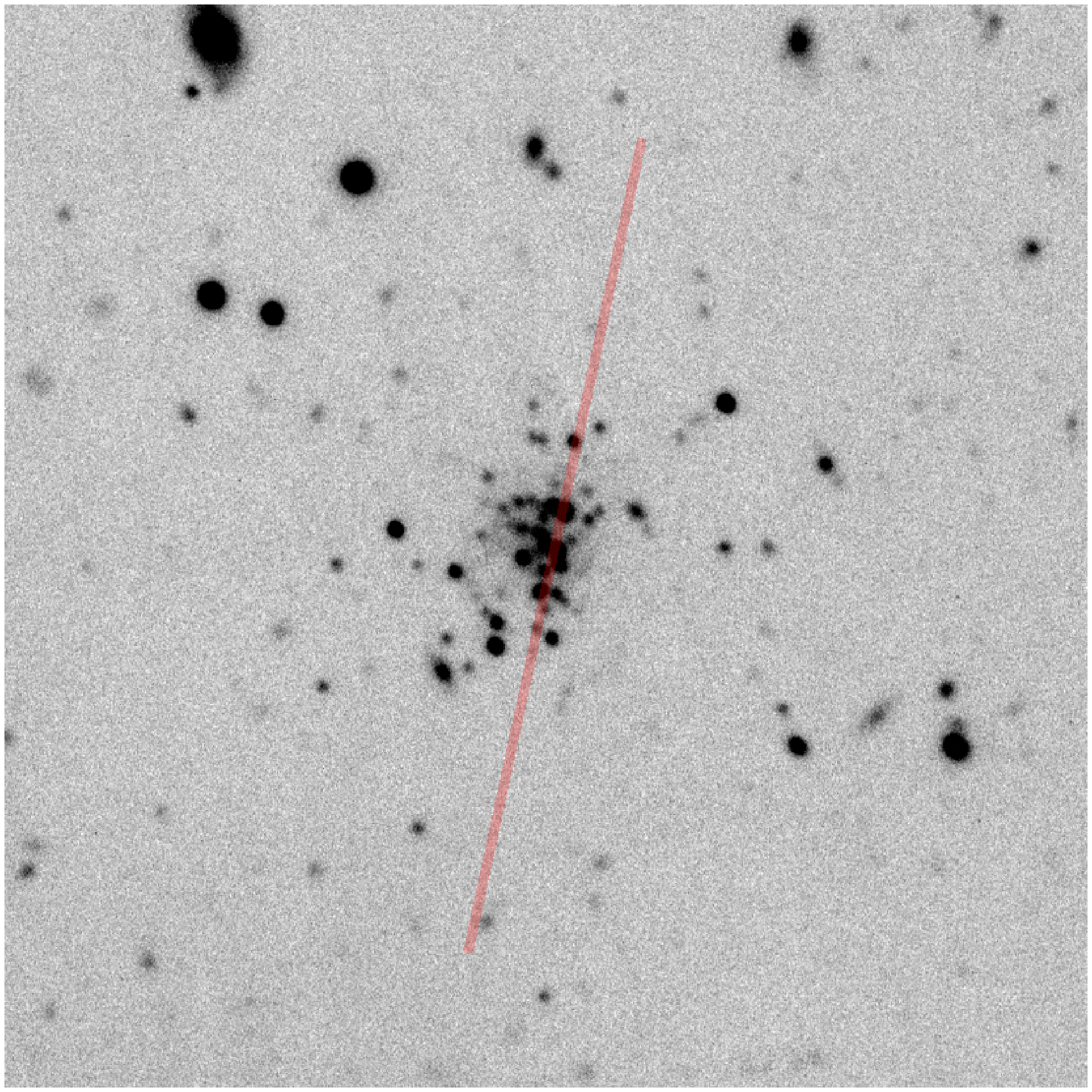}
\hspace{1mm}
\includegraphics[width=70mm]{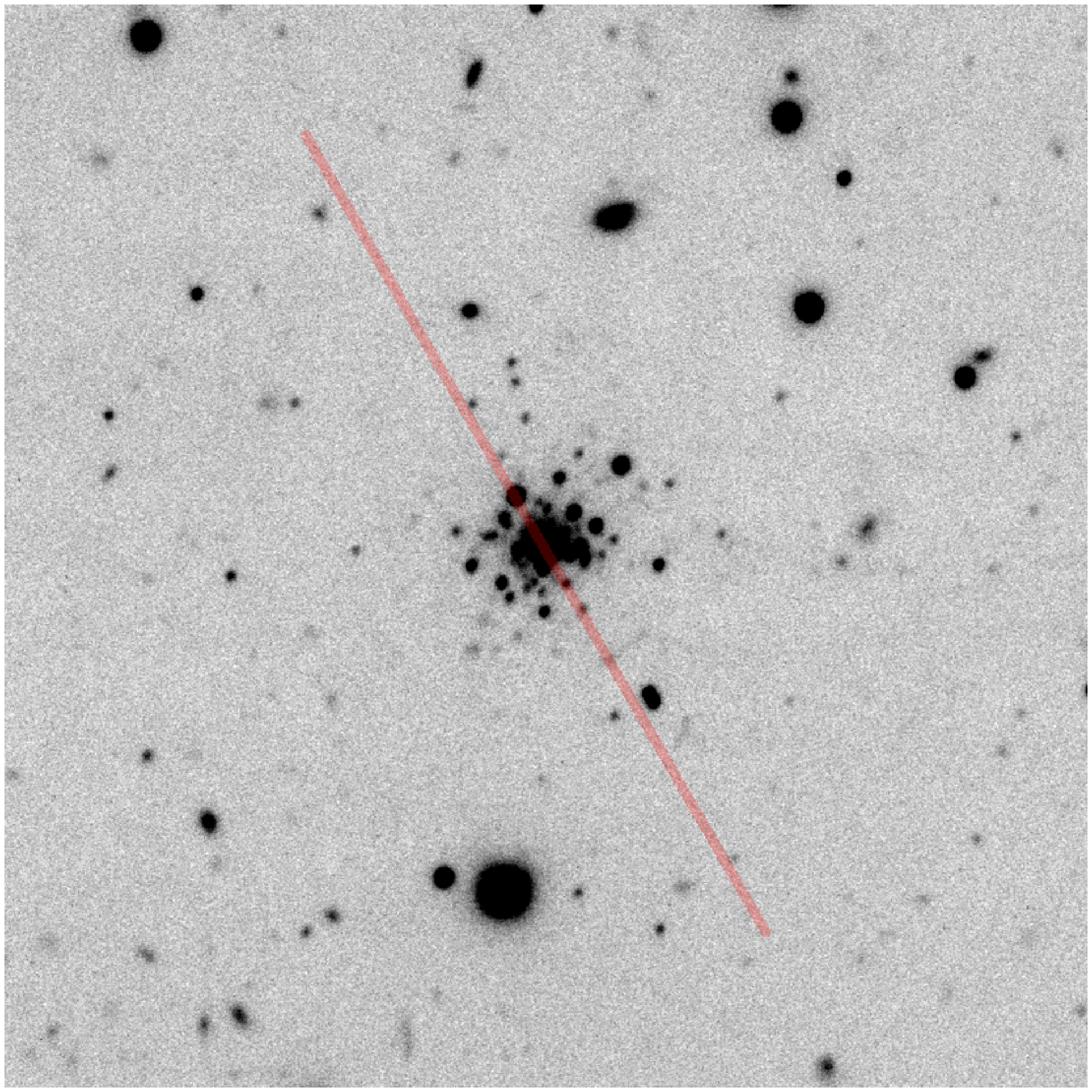}
\caption{Central $1\arcmin \times 1\arcmin$ thumbnails from our combined $2400$s $i\arcmin$-band 
GMOS images of PAndAS-7 (left) and PAndAS-8 (right). The image quality is $0.57\arcsec$ and $0.45\arcsec$, 
respectively. North is up and east is to the left. Longslit positioning and orientation is indicated.}
\label{f:images}
\end{center}
\end{minipage}
\end{figure*}

We reduced our data using the GMOS software package in {\sc iraf}. Appropriate bias
and flat-field images obtained as part of the standard GMOS baseline calibrations were downloaded 
from the Gemini science archive and then applied to each exposure using the {\sc gireduce} task. 
The three CCD frames in a given exposure were next mosaicked into a single frame using the 
{\sc gmosaic} task, and the five or six mosaicked frames for a given filter were then combined into
a single image using the {\sc imcoadd} task. 

Fig. \ref{f:images} shows the central $1\arcmin \times 1\arcmin$ regions of our combined 
$2400$s $i\arcmin$-band images of the two clusters. Stellar objects in the image of PA-7 have
a FWHM\ $\approx 7.8$ pixels, or $0.57\arcsec$, while those in the image of PA-8 have a
FWHM\ $\approx 6.2$ pixels, or $0.45\arcsec$. Although crowding precludes precise photometry
in the innermost regions of both clusters, each has a sufficient number of individual
members surrounding its centre to produce a useable colour-magnitude diagram (CMD).

We performed photometric measurements on our combined images following a procedure identical
to that which we previously applied to GMOS imaging of the exceptionally remote M31 globular cluster 
MGC1 \citep{mackey:10b}. In brief, we fit a point-spread function (PSF) model to each combined
image using the stand-alone versions of {\sc daophot ii} and {\sc allstar ii} \citep{stetson:87}. 
For any given image we began by conducting a first pass of object detection with {\sc daophot}, 
and selected $\sim 75$ relatively bright, isolated stars from the list of detections to construct an 
initial PSF. Previous experience with MGC1, along with some additional experimentation on our
present set of observations, showed that a `penny2' PSF model (Gaussian core plus Lorentzian wings)
varying quadratically across the GMOS field of view provided excellent results. We iteratively removed 
from the PSF list any stars with an error in their fit of more than three times the average, redefining 
the model each time. 

After reaching convergence (at which point there were still typically $\ga 60$ stars defining the PSF) 
we used {\sc allstar} to subtract from the image any stars neighbouring those in the PSF list. We then 
used the now completely isolated PSF stars on this subtracted image to recalculate and further refine 
the PSF model. Next, we applied this model to the original image using {\sc allstar} and subtracted 
all known stars. This subtracted image was then run back through {\sc daophot} in order to find 
faint objects missed in the first detection pass. Finally, we took the original image and our final PSF 
model and used {\sc allstar} to perform photometric measurements on the complete list of detected 
objects.

In order to eliminate non-stellar objects and stars with poor photometry, we passed the resulting
list of measurements through several quality filters based on parameters calculated by {\sc allstar}. 
Specifically, we filtered objects by the $\chi^2$ of their PSF fit, their estimated photometric error, 
and their measured sharpness relative to the PSF model. Both clusters lie at moderately low
Galactic latitude ($b \sim -20\degr$), resulting in plenty of foreground stars scattered across both GMOS 
fields that allowed us to define, in each filter, suitable ranges in these parameters 
as a function of magnitude for well-measured point sources.

To calibrate our photometry, we decided to depart from the procedure we applied to our previous
observations of MGC1. Instead of using the baseline Gemini calibration imaging of one standard
field (containing $\sim 5$ stars) per cluster per night, we cross-matched our list of stellar detections 
against the PAndAS catalogue in order to derive suitable transformations from instrumental Gemini/GMOS 
magnitudes onto the CFHT/MegaCam photometric system. We excised circular regions of radius 
$15\arcsec$ about each cluster to avoid problems associated with crowding; across the PA-7 field
we matched $131$ stars while in the PA-8 field we matched $109$ stars. This large number of 
stars, covering a wide range of colours $0.0 \la (g-i)_{\rm CFHT} \la 4.0$, allowed for a more robust 
transformation than we were able to achieve for MGC1 by employing four standard stars to calibrate 
our photometry onto the SDSS system.

\begin{figure*}
\begin{minipage}{175mm}
\begin{center}
\includegraphics[width=60mm]{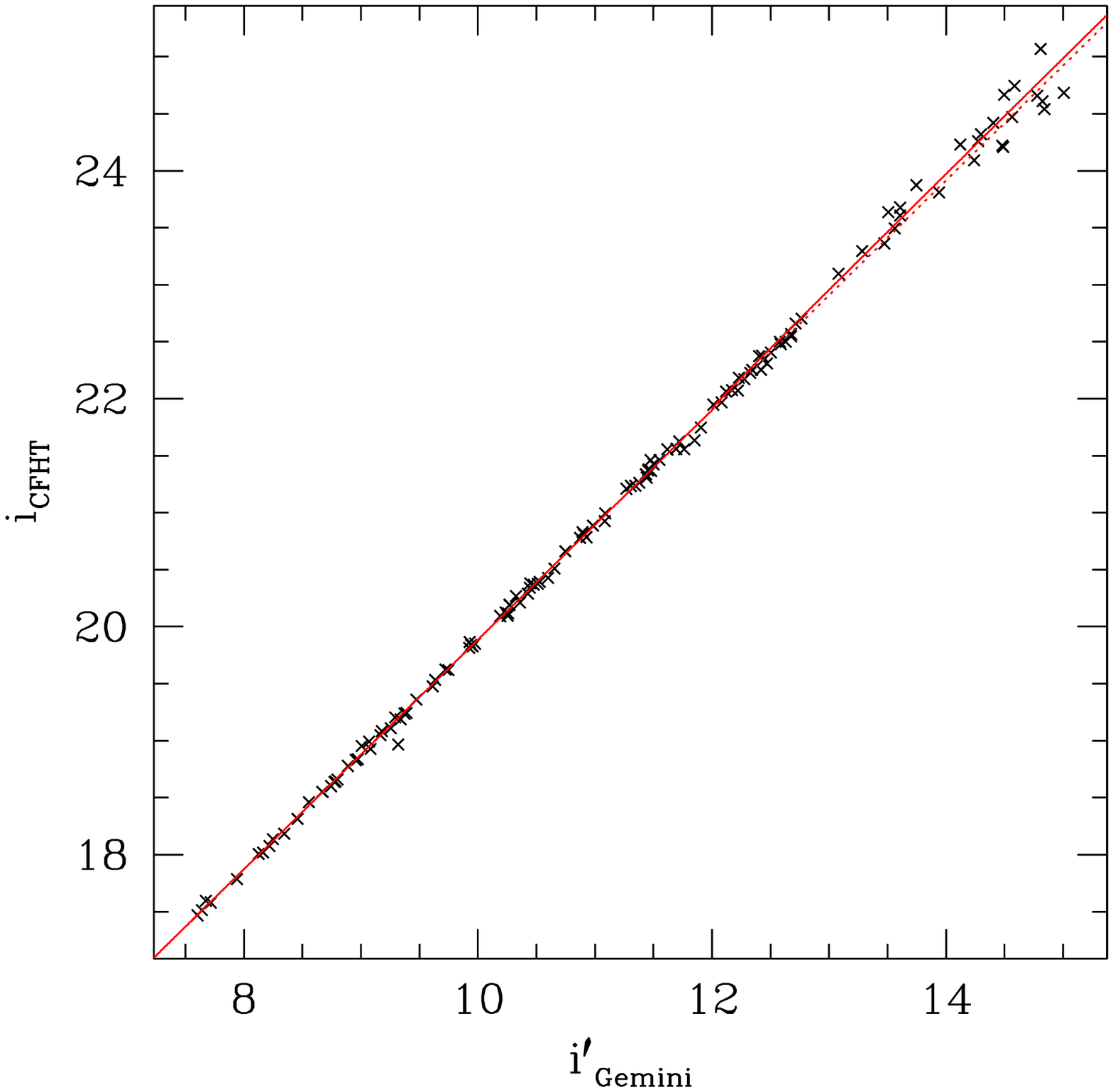}
\hspace{-2mm}
\includegraphics[width=57mm]{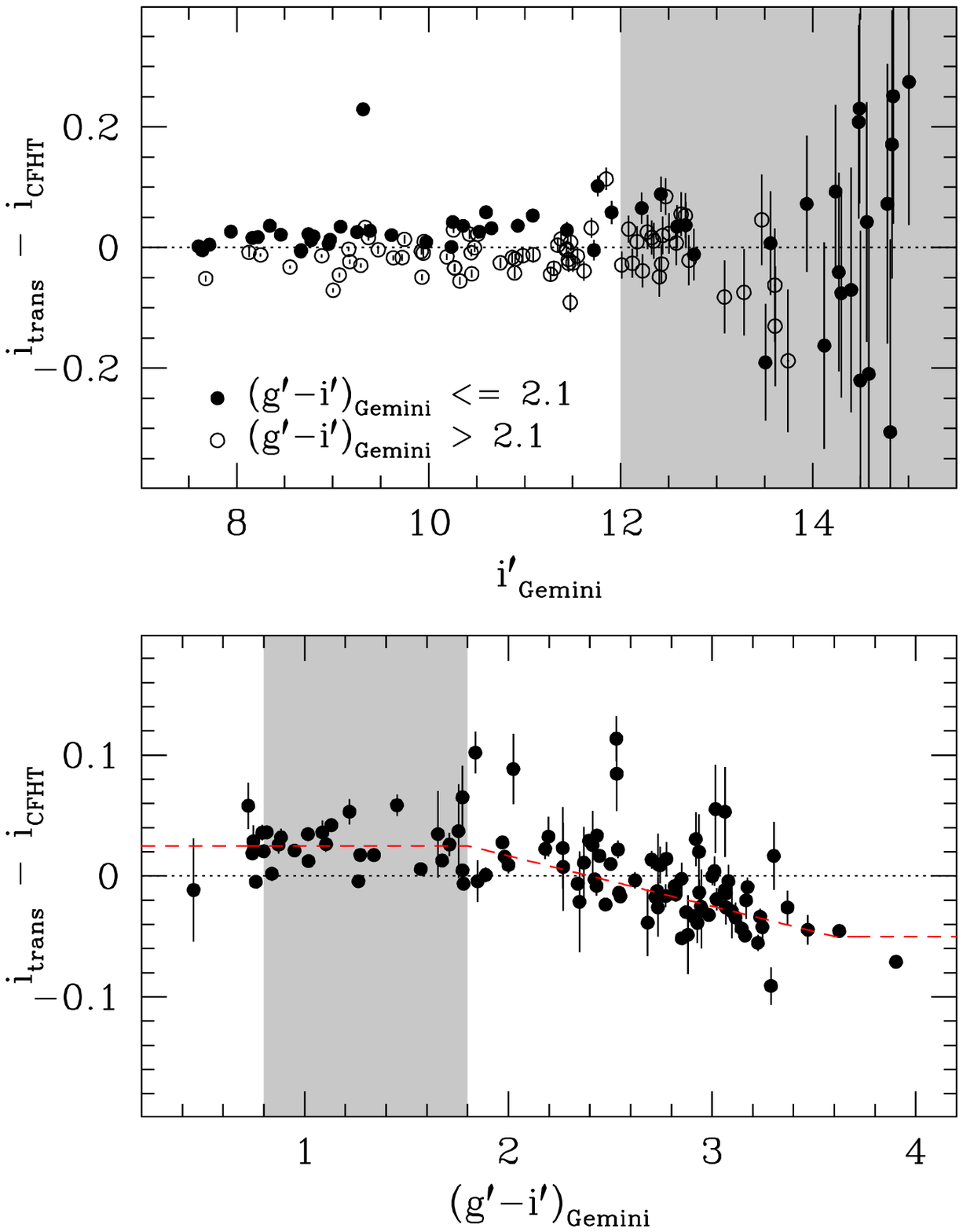}
\hspace{-2mm}
\includegraphics[width=57mm]{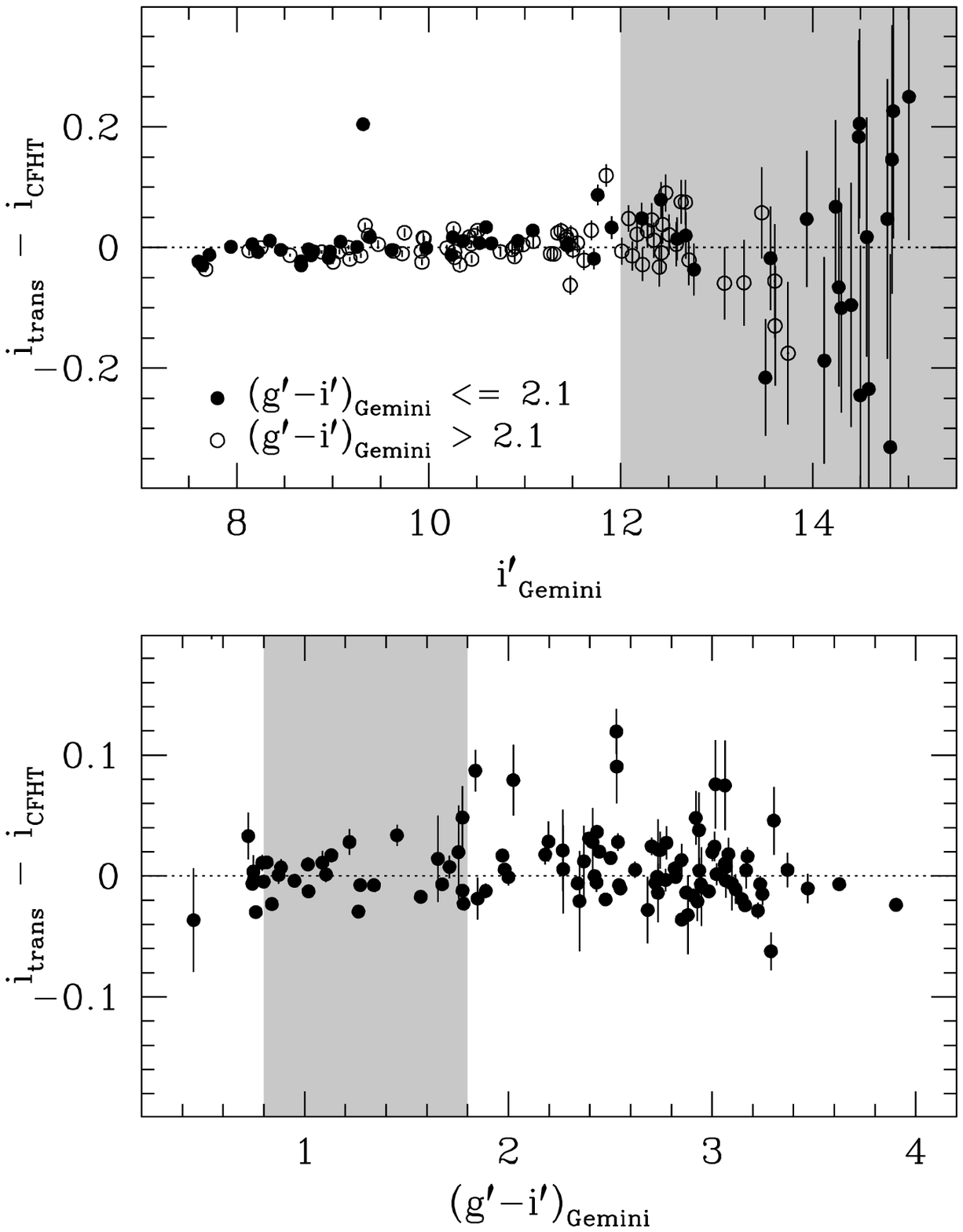}
\caption{An example of the photometric transformation from instrumental Gemini/GMOS magnitudes to the calibrated (PAndAS) CFHT/MegaCam photometric system, for PA-7 $i$-band observations. The leftmost panel shows the relationship between $i\arcmin_{\rm Gemini}$ and $i_{\rm CFHT}$. The transformation is nearly linear, except for a small exponential term that is required for faint detections. The solid line represents the linear$\,+\,$exponential relationship; the dotted line is the linear relationship only. The middle panels show the resulting residuals as a function of the instrumental $i\arcmin_{\rm Gemini}$ magnitude and $(g\arcmin - i\arcmin)_{\rm Gemini}$ colour. In the upper of these panels the solid points are for the blue half of the detections and the open points for the red half; in the lower panel for clarity we plot only detections possessing uncertainties on the residuals $\la 0.05$ mag. Both panels show that a small colour term is required -- we employ a simple prescription as marked (dashed line). The rightmost panels show the residuals after the full transformation has been applied; these are centred on zero with rms of a few$\, \times 0.01$ mag. In the central and rightmost panels we shade the region spanned by the cluster CMDs -- although to avoid crowding uncertainties we excised most cluster members from the matched detection lists before deriving the transformations.\label{f:phot}}
\end{center}
\end{minipage}
\end{figure*}

\begin{table*}
\centering
\caption{Coefficients for transforming our instrumental Gemini/GMOS photometry into the CFHT/MegaCam system.}
\begin{tabular}{@{}ccccccccccccc}
\hline \hline
Cluster and & \hspace{2mm} & \multicolumn{2}{c}{Linear Term} & \hspace{2mm} & \multicolumn{3}{c}{Exponential Term} & \hspace{2mm} & \multicolumn{4}{c}{Colour Term} \\
Passband  & & $\alpha$ & $\beta$ & & $\gamma$ & $\delta$ & $\epsilon$ & & $\kappa_1$ & $\kappa_2$ & $\kappa_3$ & $\kappa_4$ \\
\hline
PAndAS-7 $g$-band & & 1.0081 & 9.5523 & & 0.0214 & 13.0 & 1.1072 & & 0.000 & 0.000 & $...$ & $...$ \\
PAndAS-7 $i$-band & & 1.0073 & 9.8127 & & 0.0238 & 12.0 & 0.2743 & & 0.025 & -0.050 & 1.80 & 3.60 \\
\hline
PAndAS-8 $g$-band & & 1.0080 & 9.5789 & & 0.0260 & 13.3 & 0.3417 & & -0.030 & 0.020 & 1.00 & 1.80 \\
PAndAS-8 $i$-band & & 1.0001 & 9.7305 & & 0.0338 & 12.0 & 0.4752 & & 0.025 & -0.020 & 1.60 & 2.80 \\
\hline
\label{t:phot}
\end{tabular}
\end{table*}

An example of our transformation procedure is shown in Figure \ref{f:phot}. As noted previously, the 
GMOS $g\arcmin$ and $i\arcmin$ filters are similar, but not identical, to the $g\arcmin$ and $i\arcmin$ filters 
employed by CFHT. To a first approximation, we found both the $g$-band and $i$-band transformations to be linear. 
However, some experimentation revealed a mild deviation at the faint end for both filters, which we accounted for 
by adding a small exponential term at faint magnitudes. The necessity of this additional term is not evident to the 
eye in Figure \ref{f:phot}, but was readily identified after, for example, trying to match globular cluster fiducial
sequences to the colour-magnitude diagrams for PA-7 and PA-8 (Section \ref{ss:agemet}).

As in Figure \ref{f:phot}, plots of the resulting residuals revealed that further small corrections involving 
$(g\arcmin-i\arcmin)_{\rm Gemini}$ were required. The implied form of these was consistent across the $g$-
and $i$-band photometry for both clusters -- a constant offset at blue colours, moving linearly to another
constant offset at red colours.

In summary, our final transformations took the form:
\begin{equation}
m_{\rm trans} = \alpha m\arcmin + \beta + \eta(m\arcmin) - \chi(c\arcmin)\,\, .
\end{equation}
Here, $m\arcmin$ is the instrumental GMOS $g\arcmin$ or $i\arcmin$ magnitude and $c\arcmin$ is the instrumental
$(g\arcmin - i\arcmin)$ colour, $m_{\rm trans}$ is the final transformed magnitude on the calibrated (PAndAS) 
CFHT/MegaCam scale, $\alpha$ and $\beta$ are coefficients defining the linear term, and $\eta(m\arcmin)$ 
and $\chi(c\arcmin)$ are the exponential and colour terms, respectively:
\begin{equation*}
\eta(m\arcmin) = 
\begin{cases} 0 & \text{if $m\arcmin < \delta$}
\\
\gamma (m\arcmin + \delta)^{\epsilon}  & \text{if $m\arcmin \ge \delta$} 
\end{cases}
\end{equation*}
\begin{equation*}
\chi(c\arcmin) = 
\begin{cases} \kappa_1 & \text{if $c\arcmin < \kappa_3$}
\\
\frac{(\kappa_2 - \kappa_1)}{(\kappa_4 - \kappa_3)} (c\arcmin - \kappa_3) + \kappa_1  & \text{if $\kappa_3 \le c\arcmin \le \kappa_4$} 
\\
\kappa_2 & \text{if $c\arcmin > \kappa_4$} \,\, .
\end{cases} 
\end{equation*}
The coefficient $\delta$ represents the cut-on magnitude for the exponential term, while
$\kappa_3$ and $\kappa_4$ are the blue and red colours of the knees in the colour term.
We list the derived coefficients for each of our photometric transformations in Table \ref{t:phot}.
As in Figure \ref{f:phot}, application of these transformations resulted in rms residuals
at a level of a few$\,\times 0.01$ mag across the full range of colours in both filters.

The MegaCam $i$-band filter broke in June 2007 and was replaced in October 2007; these two filters are 
not identical at a level of up to $\pm 0.1$ mag across the full colour
range \citep[e.g.][]{mcconnachie:10}. The PAndAS imaging covering PA-7 and PA-8 was taken in late 
2006 and our transformations are hence to the original MegaCam $i$-band system. 

\subsection{Spectroscopy}
We obtained longslit spectra of PA-7 and PA-8 with GMOS-N on two nights in July and August 2010
with the aim of measuring radial velocities for these two faint clusters. The data were obtained in queue 
mode as part of program GN-2010B-Q-19 (PI: Mackey) during clear conditions and with seeing better than 
$0.75\arcsec$, as outlined in Table \ref{t:obs}. We employed the R831 grating operating
at central wavelengths near $8500\,$\AA, in combination with a $0.75\arcsec$ slit width and the 
RG610 blocking filter, to obtain a (measured) resolution $R \approx 3150$ around the Ca$\,${\sc ii} 
triplet at $8498\,$\AA, $8542\,$\AA, and $8662\,$\AA. We binned the CCD by a factor two in both
directions to achieve a spatial resolution of $0.1454\arcsec$\ per pixel and a dispersion of $0.68\,$\AA\ per 
pixel. PA-7 is a loosely aggregated cluster with no well defined core, so we placed the slit on the
brightest central red giant member and oriented it at a position angle of $167\degr$ east of north in order to 
cover the next brightest giant plus at least one fainter member. For PA-8 we oriented 
the slit at a position angle of $28\degr$ east of north in order to cover the unresolved cluster core and the 
brightest resolved red giant member. Positioning of the slit for both targets is indicated
in Figure \ref{f:images}.

For each cluster we obtained six $900$s exposures, split into two groups of three observed at central
wavelengths of $8450\,$\AA\ and $8500\,$\AA\ respectively. This strategy served to fill the wavelength
coverage gaps between the three GMOS CCDs as well as mitigating the effects of hot pixels and bad columns. 
Within each group the three exposures were dithered by $\pm 15\arcsec$ along the slit in order to fill the 
gaps created by the stabilising bridges on the GMOS longslit mask, further mitigate the impact of any 
detector defects, and to help reduce systematics during the subtraction of atmospheric emission lines.
We bracketed each group of three science frames with arc-lamp exposures to ensure an accurate
wavelength calibration. Flat-field images were taken for each central wavelength setting.

In addition to PA-7 and PA-8 we observed the very bright and concentrated M31 globular clusters G1
and MGC1 to serve as radial velocity templates. G1 is listed with a velocity $V_r = -332 \pm 3$\ km$\,$s$^{-1}$
in the Revised Bologna Catalogue \citep[RBC, V4.0;][]{galleti:04}, while \citet{alvesbrito:09} measured
$V_r = -355 \pm 2$\ km$\,$s$^{-1}$ for MGC1 from a high resolution spectrum. As outlined in Table \ref{t:obs},
our observations of G1 were obtained in July 2010 as part of the same program (GN-2010B-Q-19) under
which PA-7 and PA-8 were targeted, while our spectra of MGC1 were obtained in August 2011 as part of
program GN-2011B-Q-61 (PI: Mackey). For both clusters we employed an identical set-up to that described 
above for PA-7 and PA-8, except that the slit was centred on the unresolved core of each target.
The brightness of the two clusters necessitated shorter individual exposure durations of $150$s for G1
and $200$s for MGC1. We obtained nine frames for each target, split into three groups of three observed
at central wavelengths of $8450\,$\AA, $8500\,$\AA, and $8550\,$\AA. As before, within each of these
groups the three exposures were dithered by $\pm 15\arcsec$ along the slit, and each group was bracketed
with arc-lamp exposures.

\begin{figure}
\begin{center}
\includegraphics[width=80mm]{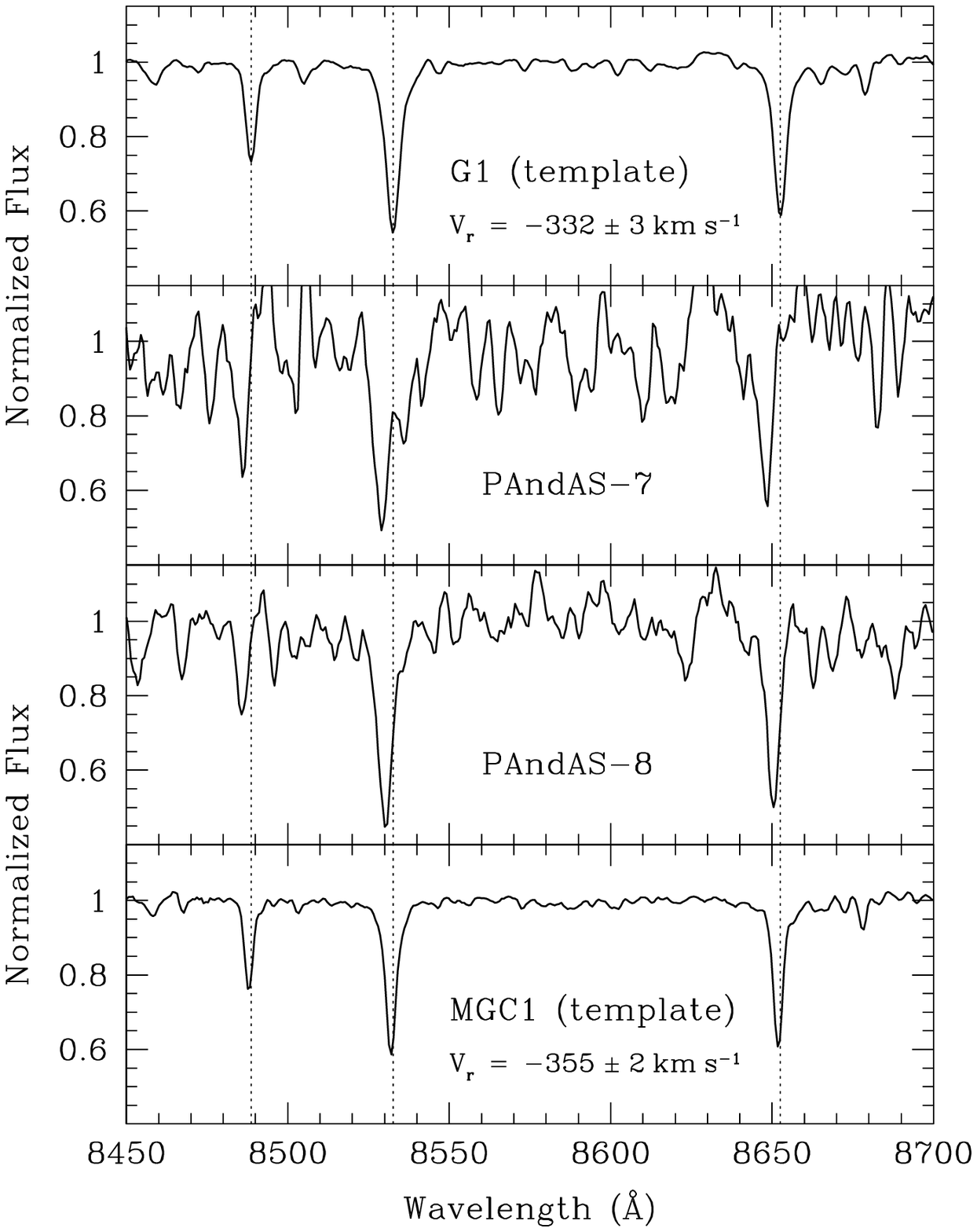}
\caption{Combined normalized spectra around the Ca{\sc ii} triplet for PA-7 and PA-8, bracketed by our two velocity template clusters G1 and MGC1. The vertical dotted lines indicate the positions of the Ca{\sc ii} lines in the spectrum for G1. Note that PA-7 and PA-8 are clearly blue-shifted with respect to these lines.}
\label{f:spectra}
\end{center}
\end{figure}
 
As with the imaging data, we reduced our spectroscopic observations using the GMOS package in {\sc iraf}.
For a given cluster, we first created normalised flat-fields for each of the central wavelength settings
using the {\sc gsflat} task. These normalised flat-fields were then applied to the science frames, along with
an appropriate bias image from the Gemini archive, using the {\sc gsreduce} task. We also ran
{\sc gsreduce} on the arc-lamp exposures, but did not apply flat-fields to these frames. Next, we ran
{\sc gswavelength} on the reduced arcs in order to establish a wavelength calibration at all points along
the slit for any given science frame, and then applied these calibrations and rectified the science frames
with {\sc gstransform}. We carefully examined the positions of the sky lines on each frame to check that 
the calibration was accurate, particularly near the object spectrum and the Ca$\,${\sc ii} triplet.
We applied {\sc gsskysub} to perform a sky subtraction on each of the calibrated science exposures,
and then extracted 1D object spectra from the sky-subtracted frames using the {\sc gsextract} task.
This step was straightforward for the two reference clusters -- for each one we simply 
extracted the spectrum of the bright unresolved core. For PA-8 we extracted the spectrum of the unresolved 
core as well as that of the bright red giant member we also placed on the slit. For PA-7 we attempted to extract 
spectra for all three of the red giants we placed on the slit; however only the brightest had sufficient signal to 
provide a useable spectrum.

Finally, we corrected each extracted spectrum to the heliocentric frame using the {\sc iraf} tasks
{\sc rvcorrect} and {\sc dopcor}, and then median-combined the individual exposures of a given
target into a final spectrum using the {\sc scombine} task. Best results, including cosmic-ray removal,
were achieved by scaling the input spectra to a common flux level and then, when combining them,
weighting each by its median flux value and applying a sigma-clipping rejection algorithm.

Figure \ref{f:spectra} shows our combined spectra in the region of the Ca$\,${\sc ii} triplet for PA-7, PA-8 
and the two reference clusters; each has been normalized using the {\sc iraf} task {\sc continuum}.
We estimate that in this part of the spectrum the signal-to-noise is $\approx 7$ per resolution element 
for PA-7, and $\approx 20$ per resolution element for PA-8. Even so, the Ca$\,${\sc ii} lines are clearly 
visible in both cases -- their positions suggest that both PA-7 and PA-8 have larger blue-shifts than 
either template cluster.

\section{Results and analysis}
\label{s:results}
\subsection{Radial velocities}
\label{ss:rv}
We used the {\sc iraf} task {\sc fxcor} to determine radial velocity estimates for PA-7 and PA-8 by 
cross-correlating their spectra against those of the template clusters G1 and MGC1 \citep[see e.g.,][]{tonry:79}. 
Because of the presence of strong telluric absorption features at several different places within our spectral 
coverage, as well as the presence of non-zero residuals from the sky subtraction, we chose to cross-correlate 
just the region covering the Ca$\,${\sc ii} triplet with a $\sim 50\,$\AA\ buffer at either end (i.e., $\sim 8450-8700\,$\AA). 
This worked well for PA-8, producing a strong isolated cross-correlation peak; however for PA-7 the results
were poor. We traced the problem for PA-7 to the Ca$\,${\sc ii} line at $8662\,$\AA\ which falls in a crowded
region of sky emission. Due to the low signal-to-noise of the source spectrum, residuals from the sky subtraction 
skewed the shape (and position) of this  Ca$\,${\sc ii} line. By restricting the cross-correlation region to the
Ca$\,${\sc ii} lines at $8498\,$\AA\ and $8542\,$\AA\ (i.e., adopting the range $\sim 8450-8570\,$\AA) we 
circumvented this problem. We verified that using just this region for PA-8, we were able to reproduce the velocity 
obtained from using all three Ca$\,${\sc ii} lines to better than the uncertainty on that measurement.

Our results are as follows: for PA-7 we obtained a heliocentric radial velocity $V_r = -433 \pm 8$\ km$\,$s$^{-1}$, 
where the uncertainty includes the quoted errors on the velocities of the reference targets. The height of the 
cross-correlation peak, against either template, was $\approx 0.5$, while the \citet{tonry:79} R-parameter 
$\approx 30$ -- both indicating a sound match between the target and reference spectra. For PA-8 we found 
$V_r = -411 \pm 4$\ km$\,$s$^{-1}$, with a cross-correlation peak of height $\approx 0.8$, and ${\rm R}\approx 49$
-- indicating a high quality match between the spectra. 

These two velocities fall quite close together and are rather different from the systemic M31 velocity of 
$-301$\ km$\,$s$^{-1}$, suggesting that PA-7 and PA-8 are likely to be associated with the underlying
South-West Cloud rather than being two unrelated members of the M31 halo. Ultimate verification of this
association will require velocity measurements for stellar members of the tidal substructure; for now we use a
simple statistical test to quantify the significance of our result. 

By measuring velocities for a large number of stars at a variety of different positions in the M31 halo, \citet{chapman:06}
estimated the (1-dimensional) halo velocity dispersion to decrease with projected radius $R$ as:
\begin{equation}
\sigma_v(R) = 152 - 0.90 \left( \frac{R}{{\rm kpc}} \right) \,\,{\rm km}\,{\rm s}^{-1}\,\, .
\end{equation}
Extrapolating a small amount beyond their outermost field at $R\approx 70$ kpc suggests that the velocity
dispersion of the M31 halo at radii comparable to PA-7 and PA-8 ($R\sim 87$ kpc) should be 
$\sigma_v(R) \approx 70$\ km$\,$s$^{-1}$. Under the assumption that remote M31 globular clusters share 
the global kinematics of the metal-poor field halo, we ask how likely it is that two independent objects, not 
members of any accreted substructure, should have velocities at least as far away
from the M31 systemic velocity as observed for both PA-7 and PA-8, but within a similar margin of each other.
We used a simple Monte Carlo model to address this question -- we generated a very large number ($\ga 10^6$)
of globular cluster pairs with velocities randomly drawn from a Gaussian distribution of width $70$\ km$\,$s$^{-1}$
centred on $V_{\rm M31} = -301$\ km$\,$s$^{-1}$, and counted the number of times that (i) both clusters
had velocities $|V_{r_{1,2}} - V_{\rm M31}| \ge 100$\ km$\,$s$^{-1}$, {\it and} (ii) the difference between them
$|V_{r_1} - V_{r_2}| \le 30$\ km$\,$s$^{-1}$. We found this to be a very unusual configuration, occuring just
$\approx 1.5\%$ of the time. Note that the bounds adopted here are somewhat conservative to allow for the
uncertainties in our radial velocity measurements, meaning that this probability is best treated as an upper limit.

This simple test indicates that PA-7 and PA-8 are almost certainly {\it not} unrelated members of the M31 halo,
consistent with our observation that these two clusters are projected directly on top of the South-West Cloud
and strongly suggestive that they are both associated with this substructure.

\begin{figure*}
\begin{minipage}{175mm}
\begin{center}
\includegraphics[width=75mm]{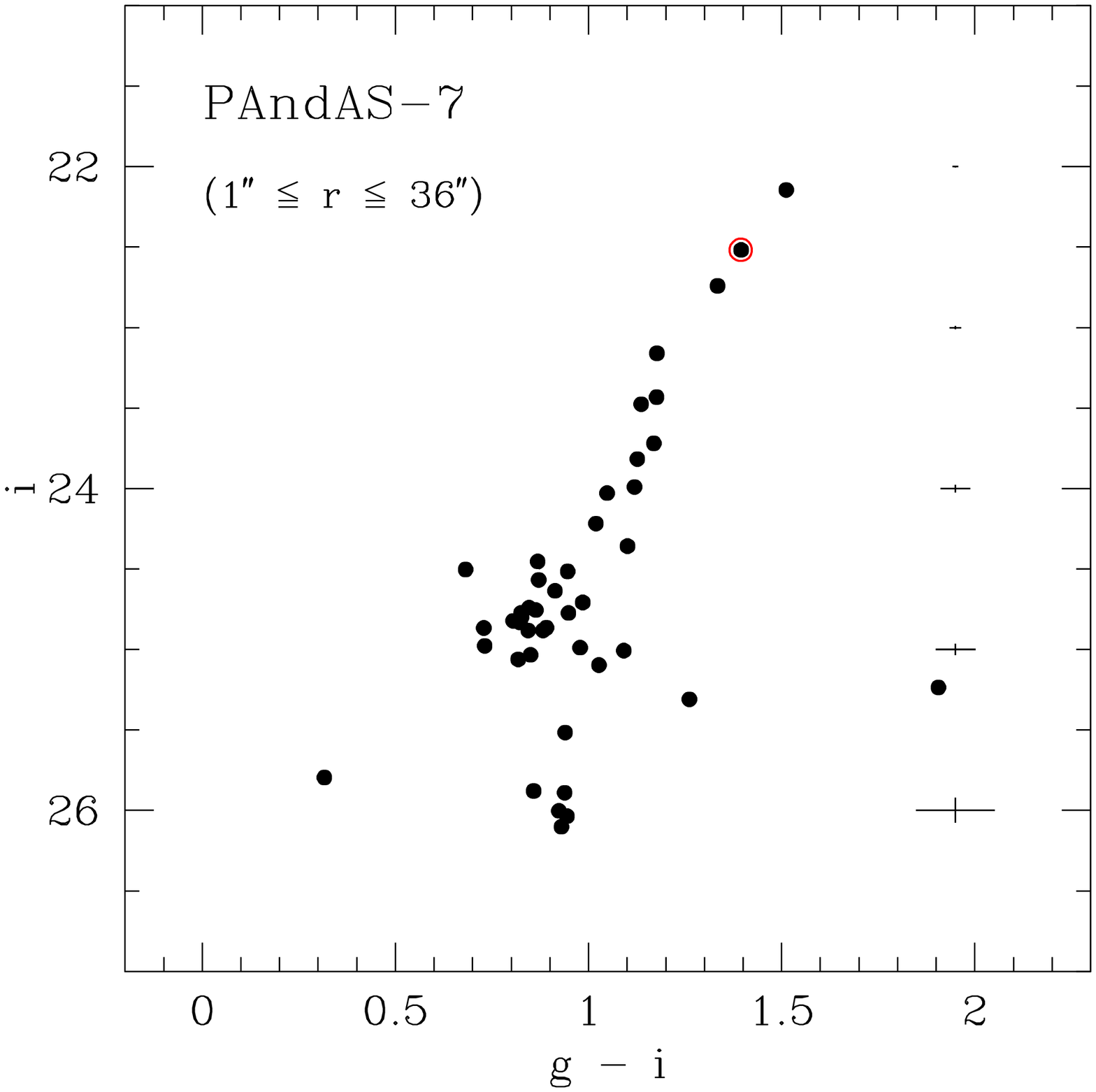}
\hspace{1mm}
\includegraphics[width=75mm]{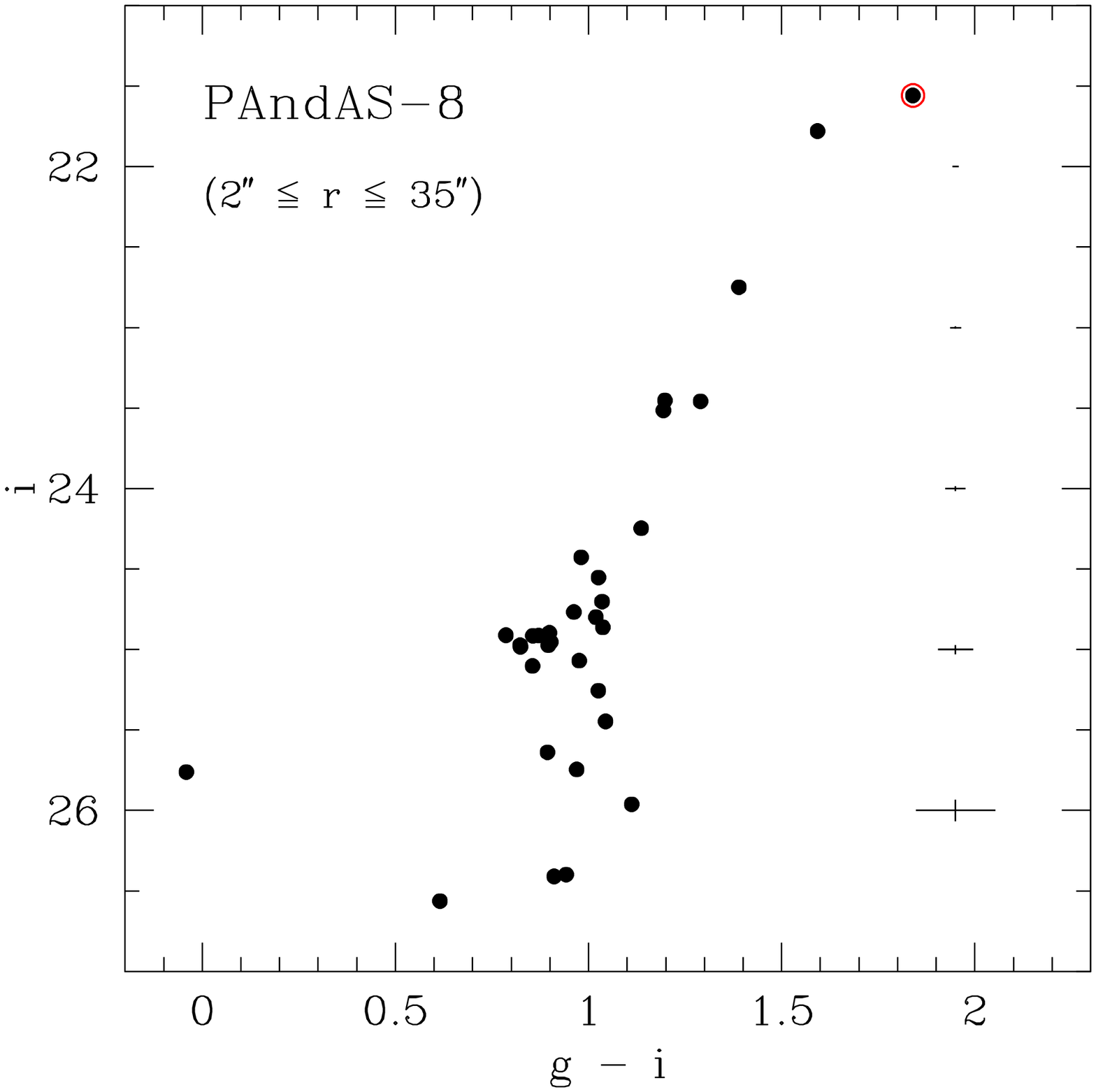}
\caption{Colour-magnitude diagrams for PA-7 and PA-8. Typical photometric uncertainties are as indicated. The upper RGB stars circled in red are those for which we obtained useful longslit spectra (note that for PA-8 we also observed the unresolved cluster core).}
\label{f:cmds}
\end{center}
\end{minipage}
\end{figure*}

\subsection{Metallicity and age constraints}
\label{ss:agemet}
Colour-magnitude diagrams for PA-7 and PA-8 are displayed in Figure \ref{f:cmds}. Our photometry reaches
to $g\approx 27$ and $i\approx 26$, just over one magnitude below the typical horizontal branch level for 
globular clusters at the M31 distance and reddening. The CMDs are sufficiently well populated to discern the 
locus of the red giant branch (RGB) and the position and morphology of the HB. For both clusters the latter feature 
is particularly striking, being very short and exclusively red -- in fact, more of a red clump than a typical globular 
cluster HB. In this respect the CMDs bear considerable resemblence to those for many of the young halo globular 
clusters found in the outskirts of the Milky Way, such as Palomar 3, 4, and 14, 
Pyxis, Eridanus, and AM-1 \citep*[e.g.,][]{stetson:99,hilker:06,dotter:08a,dotter:11}. These particular objects possess 
ages several Gyr younger than the oldest members of the Galactic system \citep[e.g.,][]{marinfranch:09,dotter:10}. 
Given this, it is quite possible that both PA-7 and PA-8 are also young globular clusters, and we assess the 
likelihood of this below.

The first point to note is that crowding and incompleteness do not affect the observed HB morphology for PA-7
and PA-8. That is, we do not believe that there are blue HB stars in these clusters that were preferentially missed 
during the photometric measurements. A by-eye inspection of the detected objects indicates that all stars at this
brightness level in the vicinity of the clusters were photometred. A few were discarded due to crowding issues
but none of these were significantly bluer than the HB stars with good measurements. As further evidence of this 
point, we direct the interested reader to our observations of the M31 halo cluster MGC1 \citep{mackey:10b}, which 
we took with GMOS under very similar atmospheric conditions. Crowding and incompleteness is more of a problem 
in this luminous compact cluster, yet we still clearly detect stars all along its blue HB.

We estimated the metallicity, foreground reddening, and distance of PA-7 and PA-8 by overplotting fiducial
sequences from several Galactic globular clusters on their CMDs. For a given comparison
cluster we measured the vertical shift required to align the HB level with that of either PA-7 or PA-8, along with 
the horizontal shift required to align the colour of the red giant branches (RGBs) at the HB level. This horizontal shift
gives the difference in colour excess between the two clusters; when combined with the vertical shift this
then yields the difference in distance modulus. Since the shape of the upper RGB is strongly sensitive to 
$[$Fe$/$H$]$, the fiducial sequence most closely tracing this feature indicates the cluster metallicity.

Because the colour of the RGB at HB level is sensitive to age as well as $[$Fe$/$H$]$ (and also, indeed, to 
$[ \alpha /$Fe$]$) we attempted, as far as possible, to adopt fiducial sequences for Galactic clusters possessing 
red HB morphology. \citet*{clem:08} provide fiducial sequences for $4$ globular clusters and one open cluster in 
the CFHT/MegaCam filter system, while \citet{an:08} provide sequences for $17$ globular clusters and $3$ open 
clusters in the SDSS filter system. Since our final photometry for PA-7 and PA-8 is calibrated to the MegaCam
system, ideally we would have used the \citet{clem:08} clusters as our templates. Unfortunately, however, a small
amount of testing revealed that of their sample, only M71 was in any way close to PA-7 and PA-8 in terms of
$[$Fe$/$H$]$ and HB morphology. 

The sample of \citet{an:08}, on the other hand, possesses a number of red HB clusters with $[$Fe$/$H$]$ comparable
to that for PA-7 and PA-8. Of particular relevance here are the outer halo objects Palomar 3 and 4. Since the SDSS 
photometric system is slightly different to the CFHT/MegaCam system, it was necessary to apply suitable 
transformations to the \citet{an:08} fiducials before comparing them to our CMDs. We used the appropriate 
equations from the online documentation describing the MegaPipe pipeline reduction process for MegaCam 
imaging\footnote{{\scriptsize {\it http://www2.cadc-ccda.hia-iha.nrc-cnrc.gc.ca/megapipe/docs/filters.html}}}:
\begin{eqnarray}
g_{{\rm MC}} &=& g_{{\rm SDSS}} - 0.153\,(g_{{\rm SDSS}} - r_{{\rm SDSS}}) \nonumber \\
i_{{\rm MC}} &=& i_{{\rm SDSS}} - 0.085\,(r_{{\rm SDSS}} - i_{{\rm SDSS}}) \label{e:sdss2mc} 
\end{eqnarray}
We also desired a more metal-rich template cluster than Pal 3 or 4. In the \citet{an:08} sample, only M71 fits
the bill; however employing this cluster as a template was a complex procedure. M71 is relatively nearby, 
meaning that its more luminous RGB members are saturated in SDSS images. The \citet{an:08} fiducial is 
consequently truncated at a level just brighter than the HB. To circumvent this issue we adopted the M71 
sequence from \citet{clem:08}. Unfortunately, these authors do not provide HB fiducials for their clusters, 
or a public archive of their photometry. Therefore, we had to merge the An et al.\ HB photometry onto
the Clem et al.\ RGB fiducial. In principle this simply requires the straightforward use of Eq. \ref{e:sdss2mc}
as in \citet{mackey:10b}; however An et al., when comparing their photometry to that of Clem et al., 
found a significant disagreement for M71 (note that all other clusters matched to a level better than $\sim 2\%$).
They ascribe this discrepancy to substantial uncertainties in their zero points for the M71 SDSS imaging.
To correct for this, we tested a grid of zero-point offsets in $gri$. Each combination of offsets was applied 
to the full An et al.\ M71 fiducial, which was then transformed using Eq. \ref{e:sdss2mc}. We located
the set of offsets producing the best match between this transformed sequence and that of Clem et al., 
and then applied this to the An et al.\ HB photometry to properly transform it onto the Clem et al.\ M71 scale.

\begin{figure*}
\begin{minipage}{175mm}
\begin{center}
\includegraphics[width=75mm]{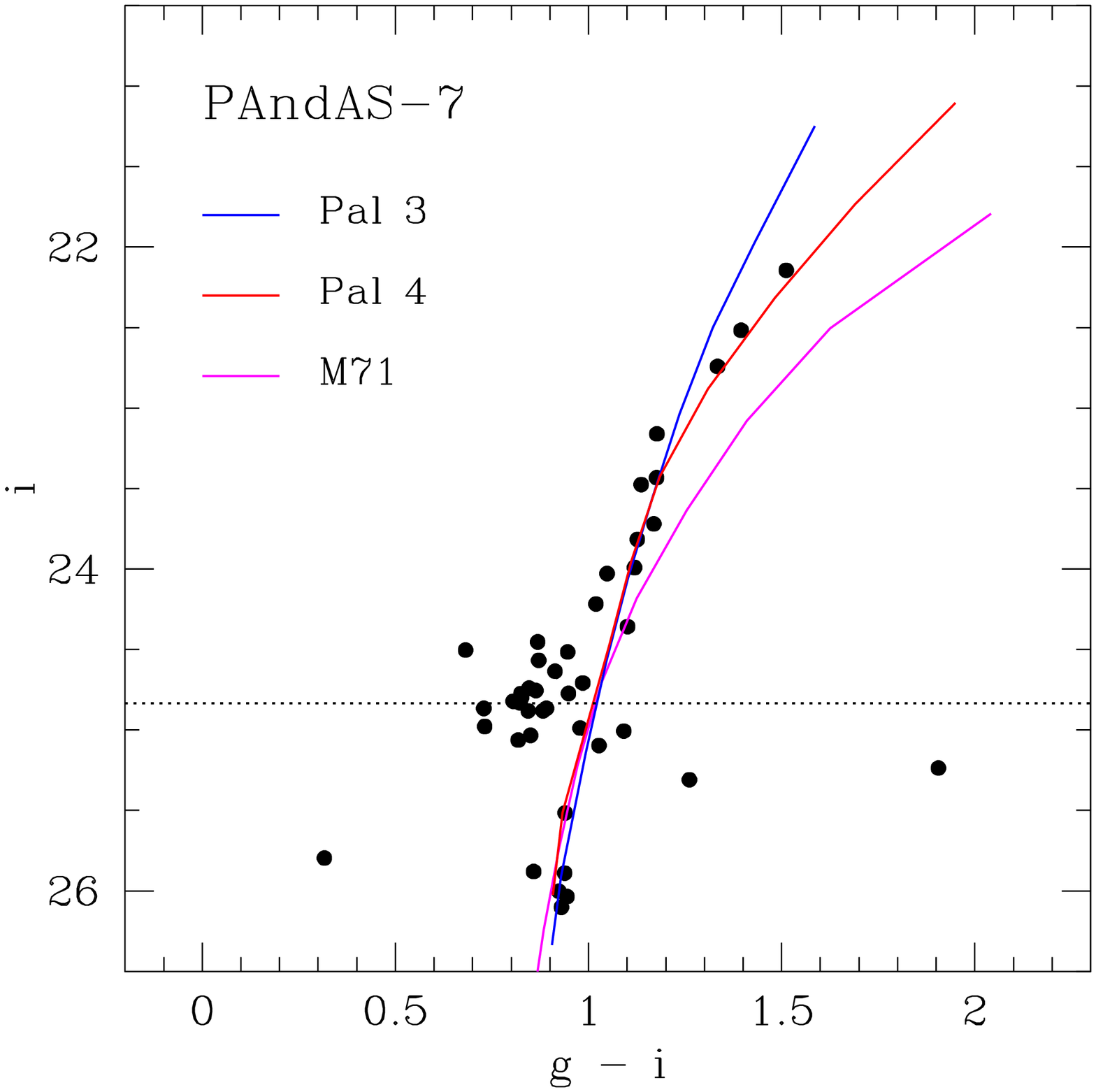}
\hspace{1mm}
\includegraphics[width=75mm]{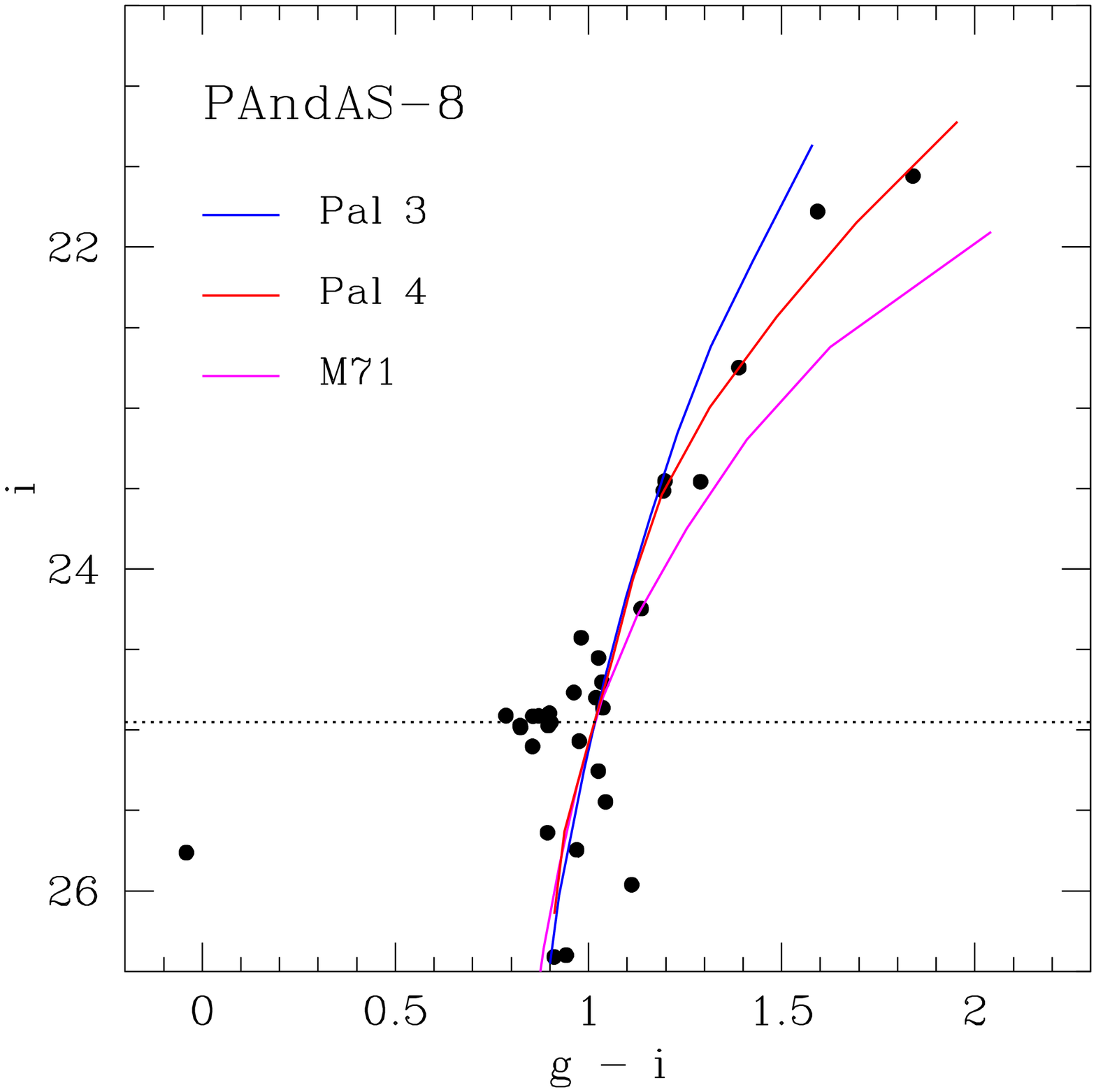}\\
\vspace{-1mm}
\includegraphics[width=75mm]{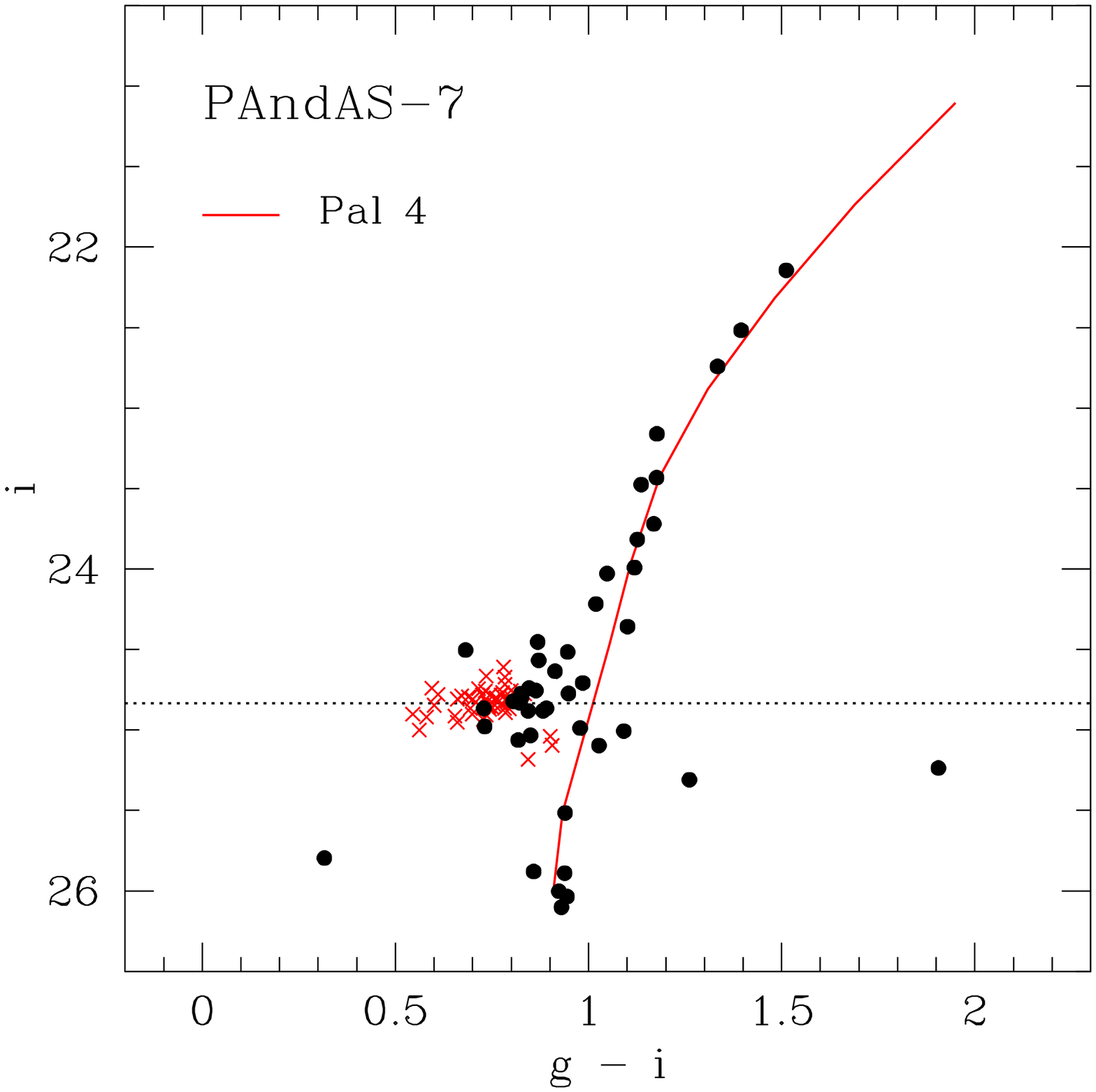}
\hspace{1mm}
\includegraphics[width=75mm]{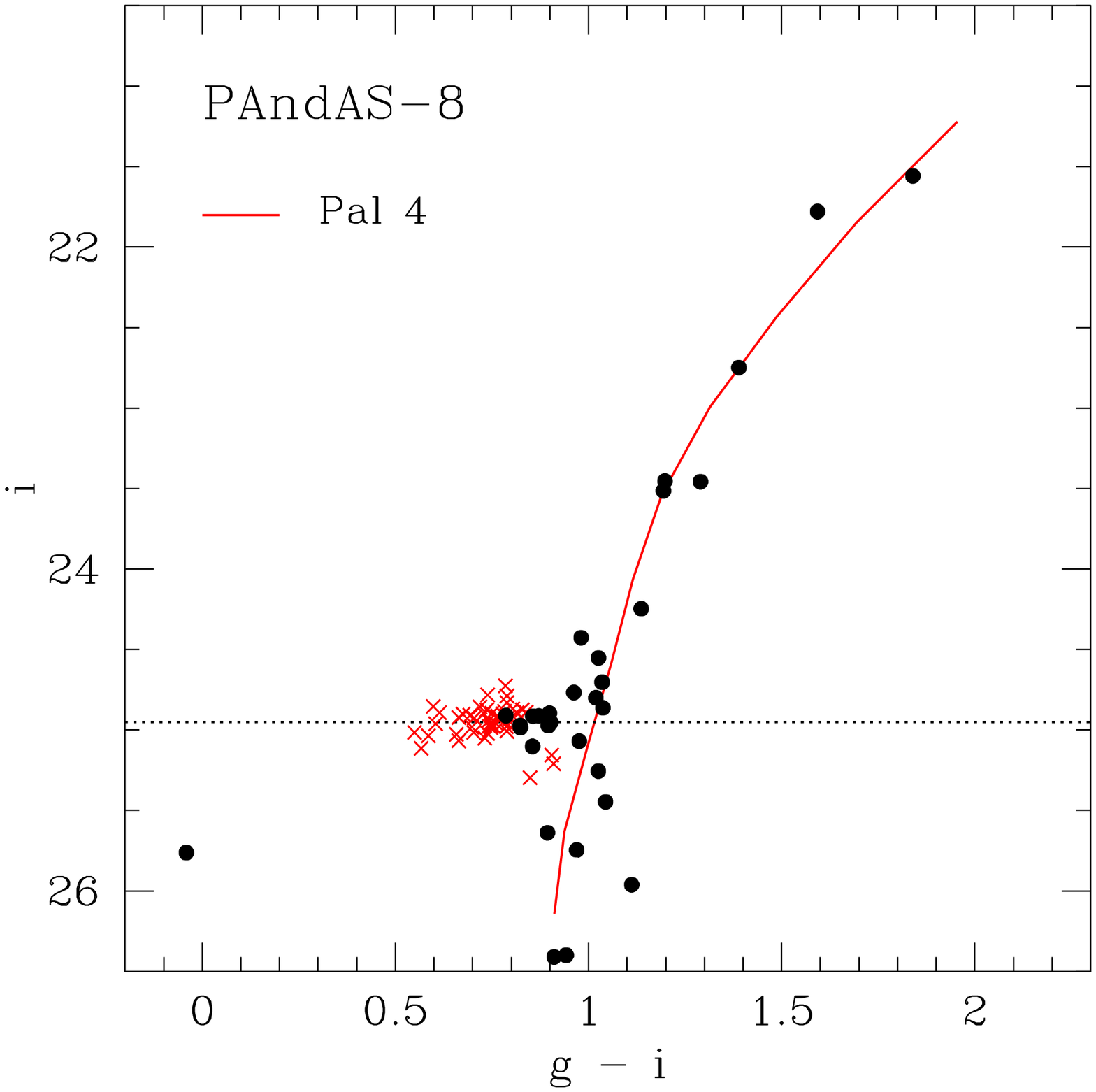}
\caption{Galactic globular cluster fiducial sequences overplotted on the CMDs for PAndAS-7 and PAndAS-8. These have been aligned using the level of the HB and the colour of the RGB at the HB level, as described in the text. The upper panels show results for three fiducial sequences of different metallicity: Palomar 3 with $[$Fe$/$H$] = -1.6$, Palomar 4 with $[$Fe$/$H$] = -1.35$, and M71 with $[$Fe$/$H$] = -0.8$. The lower panels show the fiducial sequence and HB for Palomar 4 only, which provides a good fit to the PA-7 and PA-8 CMDs. Note that in both cases the Pal 4 HB extends further to the blue and has a bluer median colour than the M31 cluster.}
\label{f:fiducials}
\end{center}
\end{minipage}
\end{figure*}
  
The results of our template alignment are presented in the upper panels of Figure \ref{f:fiducials}.
Based on the shape of the upper RGB, Palomar 3 is clearly more metal-poor than our M31 objects,
while M71 is considerably more metal-rich. The 2011 update of the \citet{harris:96} globular cluster
catalogue lists $[$Fe$/$H$] \approx -1.6$ and $-0.8$ for these two objects, respectively.
Palomar 4, on the other hand, provides an excellent upper RGB fit for both PA-7 and PA-8,
indicating that they both share a similar metallicity to this globular cluster.
The Harris catalogue lists $[$Fe$/$H$] \approx -1.4$ for Pal 4, commensurate with the value of
$[$Fe$/$H$] = -1.41 \pm 0.04$\ (statistical) $\pm 0.17$\ (systematic) measured from high
resolution spectroscopy by \citet{koch:10}. However, other publications suggest it may be slightly 
more metal-rich than this -- for example, the calcium triplet work of \citet*{armandroff:92} which
returned a value $[$Fe$/$H$] = -1.28 \pm 0.20$, or photometric studies from HST imaging
which find $[$Fe$/$H$] \approx -1.3$ \citep[e.g.,][]{stetson:99,dotter:11}. Here we adopt
$[$Fe$/$H$] = -1.35$ for Pal 4, and hence PA-7 and PA-8, but recognize a likely systematic
uncertainty of $\pm 0.1$ dex in this quantity. Based on the quality of our photometry and the
separation between the fiducial sequences of our template clusters, we estimate a statistical
uncertainty of $\pm 0.15$ dex on our best-fit metallicity for PA-7 and PA-8.

\begin{figure*}
\begin{minipage}{175mm}
\begin{center}
\includegraphics[width=150mm]{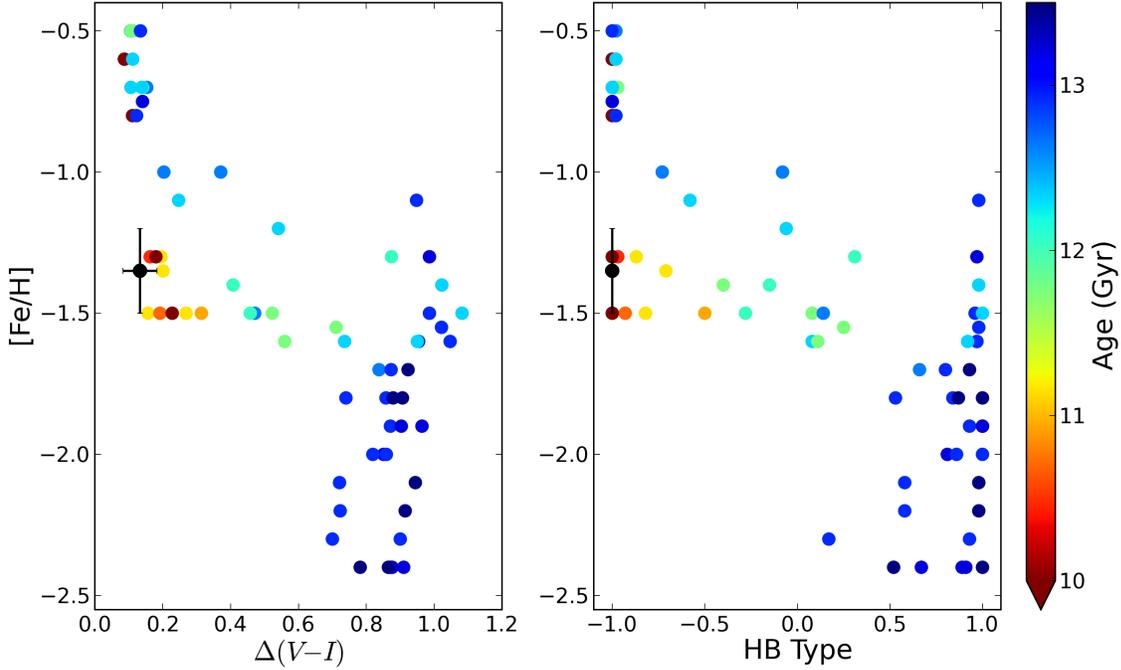}
\caption{Age as a function of metallicity and two different HB morphology metrics for a sample of $\approx 70$ Galactic globular clusters, after \citet{dotter:10,dotter:11}. The location of PA-7 and PA-8 is indicated with a black point.}
\label{f:hbmet}
\end{center}
\end{minipage}
\end{figure*}

The Harris catalogue lists a small colour excess $E(B-V) = 0.01$ for Pal 4, and an intrinsic distance
modulus $(m-M)_0 = 20.18$. Based on the vertical and horizontal shifts required to align the
template fiducial with our photometry, the implied colour excess and distance modulus for PA-7
are $E(B-V) = 0.11 \pm 0.02$ and $(m-M)_0 = 24.47 \pm 0.07$. For PA-8 we find 
$E(B-V) = 0.12 \pm 0.02$ and $(m-M)_0 = 24.58 \pm 0.07$. Note that in deriving these quantities 
we have employed the extinction coefficients from \citet{stoughton:02}, specifically that 
$A_i = 2.086\,E(B-V)$. Our quoted uncertainties reflect only the precision with which we are able to 
align the fiducial sequences with our photometry -- i.e., the uncertainties are obtained directly from 
those associated with the measured horizontal and vertical shifts. 

Our measured colour excesses are in good agreement with predictions from the reddening maps 
of \citet*{schlegel:98}: $E(B-V) = 0.09$ and $0.11$ for PA-7 and PA-8, respectively. Similarly, our 
distance moduli match quite well the canonical M31 value $(m-M)_0 = 24.47$ \citep[e.g.,][]{mcconnachie:05}.
It is possible, at marginal significance, that PA-8 could be up to $\sim 0.1$ mag more distant than PA-7. 
This would imply a significant depth of up to $\approx 40$ kpc to the South-West Cloud 
along the line of sight. However, we must bear in mind the $1\sigma$ random uncertainties of $0.07$ mag 
on each of the distance measurements, and that these do not include any systematic components 
due to, for example, the photometric calibrations, transformation of the Pal 4 fiducial to the MegaCam
photometric system, etc. It is therefore safe to say the distance estimates are consistent with each other,
although it would not be surprising if there were a mild line-of-sight extension to the South-West Cloud.

The distinctive HB morphology observed for our two M31 clusters provides an upper limit to their
ages. To demonstrate this we consider the recent work of \citet{dotter:10,dotter:11} who, using HST/ACS
photometry, investigated HB morphology in nearly half the known Galactic globular clusters as a function
of their other properties (in particular their ages and metal abundances). We reproduce one of the key results of this 
work \citep[Figure 11 from][]{dotter:11} in Figure \ref{f:hbmet}. This shows the relationship between a cluster's age,
HB morphology and $[$Fe$/$H$]$. HB morphology is parameterized using two different metrics. The first, 
introduced by \citet{dotter:10}, is the difference $\Delta(V-I)$ between the median colour of stars on the HB and 
the median colour of stars on the RGB at the HB level. The second is the commonly used ratio $(B-R)/(B+V+R)$
\citep*[e.g.,][]{lee:94}, where $B$ (alternatively, $R$) represents the number of HB stars lying to the blue (red) of the 
instability strip, and $V$ the number of RR Lyrae variables (i.e., the number of HB stars lying on the instability strip).

Irrespective of which HB metric is considered, it is clear that in the metallicity range $-1.6 \la [$Fe$/$H$] \la -1.0$
the morphology of the HB correlates strongly with the cluster age. The
oldest objects in the Galactic system have very blue horizontal branches; those several Gyr 
younger have very red horizontal branches; and there is a relatively smooth transition in between.
This metallicity range is the ``sweet spot'' for sensitivity to age -- in more metal-rich clusters the
HB morphology is driven predominantly by $[$Fe$/$H$]$ such that it is exclusively red irrespective of age,
while in more metal-poor clusters the correspondence with age is less clear as only very old systems
are observed with $[$Fe$/$H$] < -1.6$. Note that while there is some evidence that additional parameters
-- such as the central density of a cluster, or its helium abundance -- may also affect HB morphology, in
general these appear to be considerably less influential than both $[$Fe$/$H$]$ and age \citep[e.g.,][]{dotter:10}.

At $[$Fe$/$H$] \approx -1.35$, our two M31 clusters fall directly in the region where the HB is
most strongly sensitive to cluster age. Their very red HB morphologies indicate that they are likely to be
at least $\sim 2-3$ Gyr younger than the oldest Galactic globulars. To demonstrate this explicitly we 
derive the parameters $\Delta(V-I)$ and $(B-R)/(B+V+R)$ for both objects. The latter metric is 
straightforward -- all HB stars are red so the HB-type converges to $-1.0$. Although we do not explicitly
know the position of the instability strip on the CMD, the HB-type of Pal 4 is also $-1.0$ \citep[e.g.,][]{mackey:05}
and the lower panels of Figure \ref{f:fiducials} show that all HB stars in our two M31 clusters are redder
than the bluest stars on the Pal 4 HB.

To calculate $\Delta(V-I)$ for a given cluster, we drew a box around the HB and calculated the median colour 
of the stars within it. Note that because 
we observed only a small number of HB stars ($\approx 10$) in each system, the median colour is somewhat 
sensitive, at a level of roughly $\pm 0.03$ mag, to the precise boundaries of the isolating box. In principle 
we should have followed a similar procedure to determine the median colour of the RGB at the HB level; 
however we possessed too few RGB stars for this calculation to be reliable. Instead, we adopted the colour 
of the best-fitting (shifted) Pal 4 fiducial at the HB level, and conservatively assumed a total uncertainty of 
$\pm 0.05$ mag on the difference between this quantity and the median colour of the HB. These two
measurements gave us the quantity $\Delta(g-i)$, which we converted to $\Delta(V-I)$ using the 
transformation procedure outlined by \citet{martin:06} \citep[see also][]{ibata:07,huxor:08}. 
For PA-7 we measured $\Delta(V-I) = 0.14$ and for PA-8 we found $\Delta(V-I) = 0.12$.
Both values are smaller than that seen for Pal 4, $\Delta(V-I) = 0.183$ \citep{dotter:10}, consistent 
with the lower panels in Figure \ref{f:fiducials}.

\begin{figure*}
\begin{minipage}{175mm}
\begin{center}
\includegraphics[width=80mm]{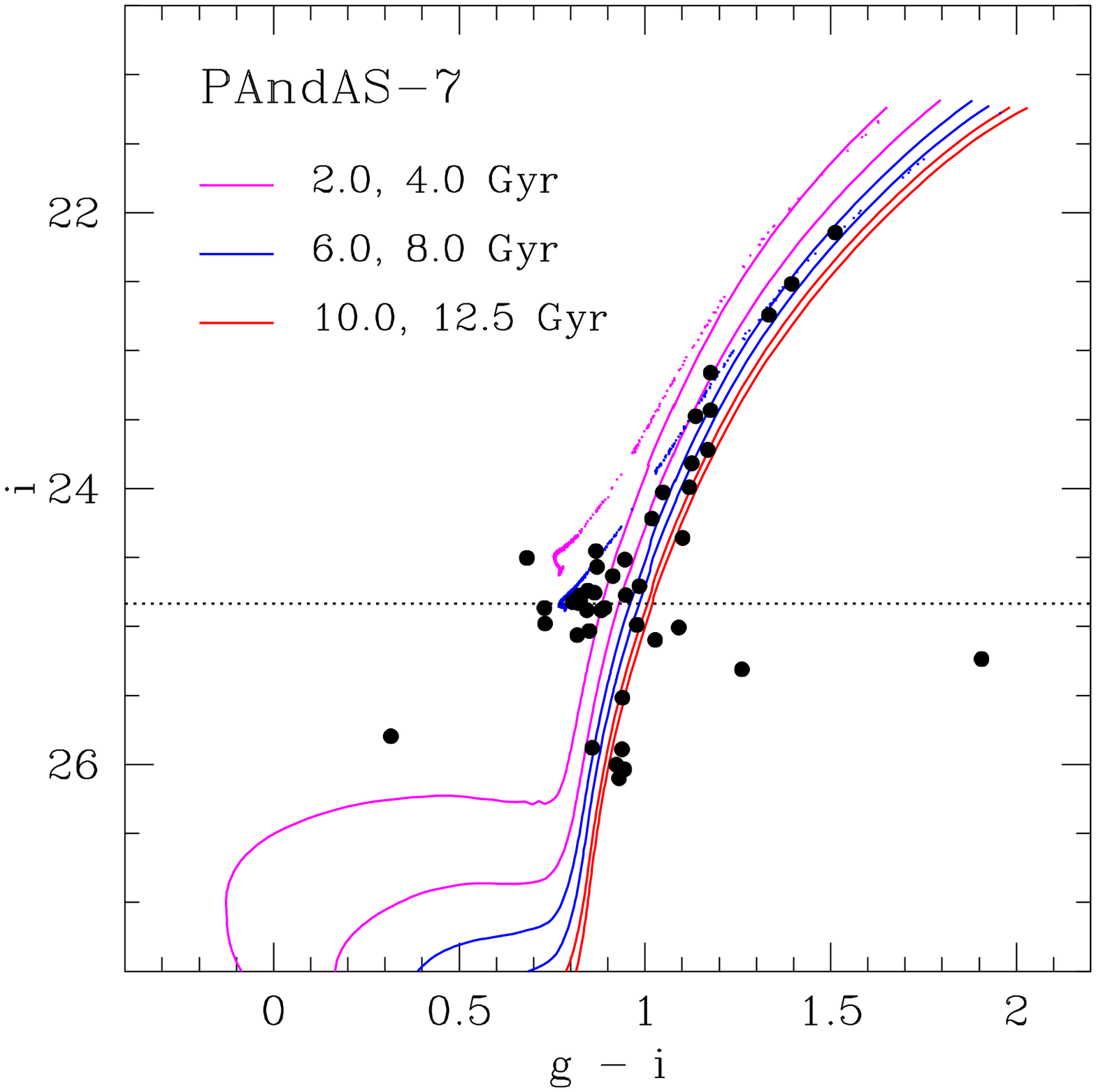}
\hspace{1mm}
\includegraphics[width=80mm]{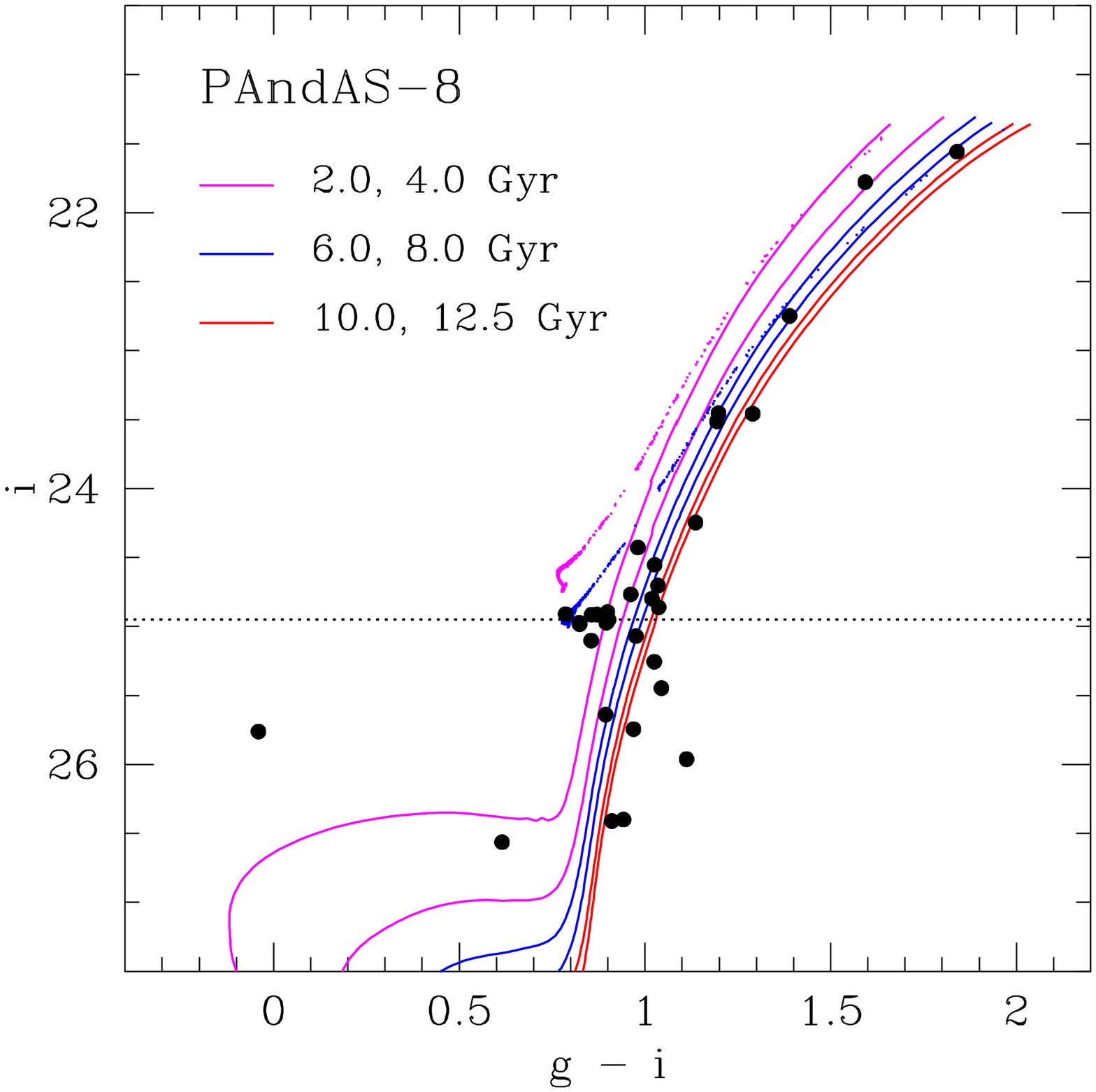}
\caption{Dartmouth isochrones and synthetic HBs overplotted on the CMDs for PA-7 and PA-8. The models have $[$Fe$/$H$] = -1.35$ and $[\alpha/$Fe$] = +0.4$, and a variety of ages as indicated. For illustrative purposes (see text) they have been shifted according to the foreground extinction levels and distance moduli derived from aligning the Palomar 4 fiducial. For clarity we do not plot a synthetic HB for the $10-12.5$ Gyr isochrones; this falls somewhat fainter ($\la 0.1$ mag) than the cluster HB levels.}
\label{f:iso}
\end{center}
\end{minipage}
\end{figure*}
 
Plotting the two HB metrics on Figure \ref{f:hbmet} confirms that PA-7 and PA-8 lie in a region inhabited 
only by Galactic globular clusters with ages {\it at least} $2$ Gyr younger than those for the oldest members 
of the Milky Way system.

While this argument provides an upper bound to the ages of PA-7 and PA-8, it says nothing about
how young these objects could be. To investigate this we fit Dartmouth isochrones and synthetic horizontal
branches \citep{dotter:08b} of different ages to our CMDs. We assume $[$Fe$/$H$] = -1.35$ and an 
$\alpha$-element abundance equivalent to that observed for Palomar 4 -- i.e., $[\alpha/$Fe$] \approx +0.4$ 
\citep{koch:10}. Note that our results are rather insensitive to the precise value adopted for 
the latter parameter.

Based on the level of the isochrone sub-giant branches, our faint detection limits of $(g,i) \sim (27.0,26.0)$ formally 
correspond to a minimum age of $\approx 2$ Gyr for PA-7 and PA-8.
However, the implied distance moduli and foreground reddening levels necessary to correctly align such young 
isochrones on the cluster CMDs suggest that their ages are, in all probability, significantly greater than this.
Figure \ref{f:iso} explicitly demonstrates this result. We plot isochrones ranging in age from
$2.0$ Gyr to $12.5$ Gyr, shifted according to the best-fit distance moduli and foreground extinction
values derived above from aligning the Palomar 4 fiducial. Because the colour of the RGB at the HB level
is mildly sensitive to age in the sense that younger clusters have somewhat bluer RGB positions,
applying these shifts does not lead to the youngest isochrones (or indeed their corresponding HB models) 
being well aligned; significantly larger values of $E(B-V)$ and $(m-M)_0$ would be necessary to
correct this. In numerical terms, aligning the $2.0$ Gyr isochrone on the PA-7 CMD implies
$E(B-V) = 0.16$ and $(m-M)_0 = 24.7$, and for PA-8 $E(B-V) = 0.17$ and $(m-M)_0 = 24.8$.
These foreground reddening levels are significantly greater than those predicted by the \citet{schlegel:98}
maps; note, however, that such a discrepancy is not out of the question \citep[see e.g.,][]{mackey:10b}.
More problematic would be the coincidence of this discrepancy with the unusually large implied distance moduli, 
which would place the clusters, respectively, $\approx 90$ and $130$ kpc behind the central disk of M31. 
Arguably a more satisfactory resolution is achieved by adopting an age in the range $\approx 6-10$ Gyr,
in which case both $E(B-V)$ and $(m-M)_0$ are much more consistent with expected values. 

Isochrones with age $12.5$ Gyr can be aligned nicely with the CMDs if, in both cases, $E(B-V)$ is about
$0.02$ mag smaller and $(m-M)_0$ about $0.1$ mag smaller than our previous best-fit values.
These are certainly within the bounds of acceptability; however as discussed previously the observed
HB morphologies of our two M31 clusters exclude such an old age (Figure \ref{f:hbmet}).

\vspace{-3mm}
\section{Discussion}
\label{s:discuss}
The measurements presented here for PA-7 and PA-8 represent a new step in improving our understanding
of the assembly of the outer halos and globular cluster systems of M31 and the Milky Way. As discussed in
Section \ref{ss:rv}, the velocities of these objects are sufficiently similar to each other, and, given their large
projected galactocentric radii, sufficiently different from the M31 systemic velocity, that it is very 
likely that they are not independent clusters but rather physically associated with the underlying South-West Cloud
(although with the caveat that to establish the link directly will require velocities for stellar members of this
overdensity). \citet{collins:09} have previously shown that the extended M31 cluster HEC12 
(also sometimes called EC4) shares a common velocity with the underlying halo feature known as Stream C. 
Apart from this, our measurements for PA-7 and PA-8 represent the strongest confirmation to date that some
remote  M31 
halo globular clusters are members of stellar substructures -- despite this commonly being assumed to be 
the case as a result of the statistically significant correlation between the spatial positions of many such 
objects and a number of the field halo substructures \citep{mackey:10a}.

In this context the resolved properties of PA-7 and PA-8 take on added significance.
We have shown that both have $[$Fe$/$H$] = -1.35 \pm 0.15$, and both possess a strikingly red horizontal branch 
morphology for this metallicity that implies their ages must be at least $\sim 2$ Gyr younger than the oldest Galactic 
globular clusters. With such characteristics
PA-7 and PA-8 would be unambiguously classified as members of the ``young halo'' subsystem in the 
Milky Way. \citet{perina:12} have recently demonstrated that many halo globular clusters in M31 exhibit an even
stronger second parameter effect than do the young halo clusters in the Milky Way. They attribute this to
the M31 objects having formed $\sim 1-2$ Gyr later than their Milky Way counterparts, which is entirely
consistent with our conclusions for PA-7 and PA-8.

As discussed in the introduction to this paper, there is a wealth of circumstantial evidence that 
the young halo population of globular clusters has been accreted into the outer Galaxy from now-defunct 
satellite systems; however, direct observations of this process in action are limited to a handful of members 
and ex-members of the Sagittarius dwarf. Our measurements provide new, independent 
evidence for close analogues of Galactic young halo clusters being accreted into the halo of a large spiral
galaxy, in a manner entirely consistent with our understanding of the origin of the outer parts of the Milky
Way globular cluster system. 

The complementary viewpoint is also interesting. If we accept {\it a priori} that the available evidence renders 
the accretion picture largely correct for the Milky Way (despite very few clusters having been directly associated 
with stellar streams), then our observations of PA-7 and PA-8 are significant in that they explicitly demonstrate 
that objects with very similar, if not identical, properties to the accreted members of the Galactic system are also 
being accreted into M31. In fact, combined with the results of \citet{mackey:10a} and \citet{perina:12}, our
results suggest that this process has occurred on a grand scale in M31.  This would add credence to the idea 
that such processes play a key role in 
determining the composition of globular cluster systems in large spiral galaxies in general.
It is of considerable interest to extend our study to include resolved observations of a more statistically robust 
number of globular clusters lying on (and off) a wide variety of M31 substructures in order to test how far the 
similarities with the outer Galactic globular cluster system stretch.

A number of authors \citep[e.g.,][]{marinfranch:09,forbes:10,dotter:10,dotter:11} have demonstrated that
the age-metallicity relation (AMR) for Galactic globular clusters exhibits two distinct loci -- one at almost
constant old age ($\approx 13-13.5$ Gyr) and a second branching to younger ages at $[$Fe$/$H$] \ga -1.5$. 
The younger branch is composed mainly of young halo clusters including several of those associated with the
Sagittarius dwarf. 
It is interesting to consider PA-7 and PA-8 in this setting. Our measured $[$Fe$/$H$] \approx -1.35$ places 
these two objects
just above the branch-point on the Galactic globular cluster AMR. At this metallicity, Galactic globular
clusters are seen with ages up to $\sim 3.5$ Gyr younger than the oldest members of the system
(see e.g., Figure \ref{f:hbmet}). As discussed in Section \ref{ss:agemet}, we formally
constrain the ages of PA-7 and PA-8 to lie in the range $\sim 2.0 - 11.0$ Gyr; however in all likelihood
they have ages older than $\approx 6$ Gyr. Thus it is quite possible that these two objects fit well with the
younger branch of the Galactic globular cluster AMR, in which case their galaxy of origin -- the progenitor
of the South-West Cloud -- should have had a cluster enrichment history similar to those satellite galaxies
which have deposited their members in the outer Milky Way. 

On the other hand, it is also possible that
PA-7 and PA-8 have younger ages than any known Galactic globular cluster at their metallicity, in which
case the implied cluster enrichment history of the SW Cloud progenitor would be comparatively rather slow, 
perhaps akin to that seen in the SMC. Note that a handful of globular clusters with ages as young as 
$\approx 6-9$ Gyr are seen in the Milky Way (e.g., Palomar 1 and 12, and Terzan 7); however these objects 
are all considerably more metal-rich than PA-7 and PA-8, with $-0.9 \la [$Fe$/$H$] \la -0.5$. At least two 
(Pal 12 and Ter 7) are associated with the Sagittarius dwarf. To more accurately constrain the ages of PA-7 
and PA-8, and thus learn something about the cluster AMR of their host galaxy, would require observations 
approaching their main-sequence turn-offs on the CMD -- which at present can only be achieved with HST.
Failing this, multicolour broadband integrated photometry could be used to place less stringent constraints
on the ages of these two objects.

\vspace{-3mm}
\section{Summary \& Conclusions}
In this paper we have presented spectroscopy and colour-magnitude diagrams for two
newly-discovered globular clusters in the remote halo of M31. These two objects, PA-7 and PA-8, project
onto a field substructure known as the South-West Cloud, leading previous authors \citep[e.g.,][]{mackey:10a}
to suggest that they have probably been accreted into the M31 system. We measure radial velocities of  
$V_r = -433 \pm 8$\ km$\,$s$^{-1}$ for PA-7 and $-411 \pm 4$\ km$\,$s$^{-1}$ for PA-8. 
These are rather similar to each other, especially given their moderately large separation from the M31 systemic 
velocity. Based on the known velocity dispersion of metal-poor stellar populations in the outer parts of the M31 halo, 
we used a simple Monte Carlo model to show that observing velocities similar to those for PA-7 and PA-8 is rather 
unlikely if these objects are completely independent, with a probability of $\la 1.5\%$. This result
reinforces the link between PA-7, PA-8 and the SW Cloud and provides strong evidence that these clusters
have indeed been accreted into the M31 halo.

Our colour-magnitude diagrams indicate that both PA-7 and PA-8 have metallicities $[$Fe$/$H$] = -1.35 \pm 0.15$.
A reference fiducial for the remote Galactic globular cluster Palomar 4 provides an excellent fit to the 
red giant branches in both clusters. From the horizontal and vertical shifts necessary to correctly align the 
Pal 4 sequence we derive foreground extinctions consistent with those predicted by the reddening maps of
\citet{schlegel:98} and distance moduli commensurate with the canonical M31 distance. PA-8 is up to $\sim 0.1$ mag
($\approx 40$ kpc) more distant than PA-7 at marginal significance; if this observation is correct then it
implies a moderate line-of-sight depth to the SW Cloud that could, in future, be used to help constrain
the geometry of the orbit of its progenitor galaxy.

The most striking aspect of the PA-7 and PA-8 colour-magnitude diagrams is the extremely short, red
horizontal branch morphology exhibited by both clusters. In the Galactic halo, clusters with similar
$[$Fe$/$H$]$ and HB morphology are exclusively $\ga 2$ Gyr younger than the oldest members of the 
system. This implies that PA-7 and PA-8 are very likely to have similarly young ages. Formally, our lower
age limit for both clusters is $\approx 2$ Gyr; however we provide circumstantial arguments that they
are probably $\ga 6$ Gyr in age. Our observations provide strong evidence for young globular clusters being 
accreted into the halo of a large spiral galaxy, in a manner entirely consistent with our picture for the 
assembly of the outer Milky Way globular cluster system. These results further add credence to the 
idea that such processes play a central role in determining the composition of globular cluster systems in 
large spiral galaxies in general.

\vspace{-3mm}
\section*{Acknowledgments}
We would like to thank Aaron Dotter for his help in constructing Figure \ref{f:hbmet}.
We appreciate suggestions from the referee which helped improve the manuscript.
ADM is grateful for support by an Australian Research Fellowship (Grant DP1093431)
from the Australian Research Council.  AMNF, ADM, and APH acknowledge support by
a Marie Curie Excellence Grant from the European Commission under contract 
MCEXT-CT-2005-025869 during which this work was initiated. GFL thanks the Australian 
Research Council for support through his Future Fellowship (FT100100268) and
Discovery Project (DP110100678).

This paper is based on observations obtained at the Gemini Observatory, which is operated 
by the Association of Universities for Research in Astronomy, Inc., under a cooperative 
agreement with the NSF on behalf of the Gemini partnership: the National Science Foundation 
(United States), the Science and Technology Facilities Council (United Kingdom), the
National Research Council (Canada), CONICYT (Chile), the Australian Research Council
(Australia), Minist\'{e}rio da Ci\^{e}ncia e Tecnologia (Brazil) and SECYT (Argentina).
These observations were obtained under programs GN-2008B-Q-22, GN-2010B-Q-19,
and GN-2011B-Q-61.

\label{lastpage}

\end{document}